\documentclass[12pt]{article}

\pdfoutput=1

\usepackage[top=80pt,bottom=85pt,left=85pt,right=85pt]{geometry}
\usepackage{amssymb}
\usepackage{amsmath}
\usepackage{setspace}
\usepackage{sectsty}
\usepackage{accents}
\usepackage{comment,cite}
\usepackage[english]{babel}

\usepackage[usenames]{xcolor}
\usepackage{graphicx,subcaption}
\usepackage{setspace}
\usepackage{float}
\usepackage{booktabs}
\usepackage[debug,pageanchor=false]{hyperref}
\definecolor{link}{rgb}{.8,.15,.1}
\definecolor{pigment}{rgb}{0.36, 0.54, 0.66}
\definecolor{pigment2}{rgb}{0.19, 0.55, 0.91}
\definecolor{pigment3}{rgb}{0.2, 0.2, 0.6}
\definecolor{light-gray}{gray}{0.75}
\hypersetup{colorlinks=true,linkcolor=link,citecolor=link,urlcolor=link,linktocpage}
\usepackage{cleveref}

\usepackage{array,multirow,booktabs,longtable}
\usepackage{mathrsfs}
\usepackage{tikz-cd} 
\usetikzlibrary{backgrounds, arrows,calc,shapes,decorations.pathreplacing, decorations.markings, automata,positioning}

\tikzset{
        cvertex/.style={circle,draw=black,inner sep=1pt,outer sep=3pt},
        vertex/.style={circle,fill=black,inner sep=1pt,outer sep=3pt},
        star/.style={circle,fill=yellow,inner sep=0.75pt,outer sep=0.75pt},
        tvertex/.style={inner sep=1pt,font=\scriptsize},
        gap/.style={inner sep=0.5pt,fill=white}}

\tikzstyle{mybox} = [draw=black, fill=blue!10, very thick,
    rectangle, rounded corners, inner sep=10pt, inner ysep=20pt]
\tikzstyle{boxtitle} =[fill=blue!50, text=white,rectangle,rounded corners]

\newcommand{\nn}{\mathbb{N}}

\newcommand{\cc}{\mathbb{C}}
\newcommand{\zz}{\mathbb{Z}}
\newcommand{\pp}{\mathbb{P}} 
\newcommand{\qq}{\mathbb{Q}} 
\newcommand{\None}{{\mathcal{N}=1}}

\newcommand*{\equal}{=}

\DeclareMathOperator{\Tr}{Tr}
\DeclareMathOperator{\SU}{SU}

\DeclareMathOperator{\U}{U}
\DeclareMathOperator{\SO}{SO}

\DeclareMathOperator{\SL}{SL}

\DeclareMathOperator{\coker}{coker}

\DeclareMathOperator{\End}{End}
\DeclareMathOperator{\Hom}{Hom}

\DeclareMathOperator{\Spec}{Spec}

\DeclareMathOperator{\diag}{diag}

\newcommand{\todo}[1]{}
\renewcommand{\todo}[1]{{\color{red} TODO: {#1}}}
\newcommand{\red}[1]{}
\renewcommand{\red}[1]{{\color{red} {#1}}}
\newcommand{\blue}[1]{}
\renewcommand{\blue}[1]{{\color{blue} {#1}}}

\makeatletter
\newcommand\xleftrightarrow[2][]{%
  \ext@arrow 9999{\longleftrightarrowfill@}{#1}{#2}}
\newcommand\longleftrightarrowfill@{%
  \arrowfill@\leftarrow\relbar\rightarrow}
\makeatother

\makeatletter
\@addtoreset{equation}{section}
\makeatother

\usepackage[en-US,showdow,showzone]{datetime2}
\DTMlangsetup[en-US]{abbr=true,showyear=false,zone=atlantic}

\begin{document}


\begin{titlepage}

\begin{center}

\vskip .3in \noindent

{\Large \bf{Holography, Matrix Factorizations and K-stability}}

\bigskip

Marco Fazzi$^{a,b}$ and Alessandro Tomasiello$^{c,d}$\\

\bigskip


\bigskip
{\small

$^a$ Department of Physics, Technion, 32000 Haifa, Israel\\
\vspace{.25cm}
$^b$ Department of Particle Physics and Astrophysics, \\Weizmann Institute of Science, 76100 Rehovot, Israel \\
\vspace{.25cm}
$^c$ Dipartimento di Fisica, Universit\`a di Milano--Bicocca, \\ Piazza della Scienza 3, I-20126 Milano, Italy \\ 
\vspace{.25cm}
$^d$ INFN, sezione di Milano--Bicocca, Piazza della Scienza 3, I-20126 Milano, Italy
	
}

\vskip .3cm
{\small \tt 
\href{mailto:mfazzi@physics.technion.ac.il}{mfazzi@physics.technion.ac.il} \hspace{.5cm} \href{mailto:alessandro.tomasiello@unimib.it}{alessandro.tomasiello@unimib.it}
}

\vskip .6cm
     	{\bf Abstract }
\vskip .1in
\end{center}

Placing D3-branes at conical Calabi--Yau threefold singularities produces many AdS$_5$/CFT$_4$ duals. Recent progress in differential geometry has produced a technique (called K-stability) to recognize which singularities admit conical Calabi--Yau metrics. On the other hand, the algebraic technique of non-commutative crepant resolutions, involving matrix factorizations, has been developed to associate a quiver to a singularity. In this paper, we put together these ideas to produce new AdS$_5$/CFT$_4$ duals, with special emphasis on non-toric singularities.


\vfill
\eject

\end{titlepage}


\tableofcontents

\section{Introduction} 
\label{sec:intro}

An important feature of string theory is that it makes sense on spaces with singularities. In particular, D-branes on such spaces can get stuck at the singular loci, giving rise to intricate algebraic structures that can be described by quiver diagrams. This plays an important role in holography: placing a stack of D3-branes at a conical singularity and taking a near-horizon limit, one obtains an AdS$_5$ solution that is dual to a CFT$_4$ described by the quiver.  

The cleanest example of this procedure is when the conical space is a Calabi--Yau (CY) threefold $Y$. In the singular case, there are several possible definitions of CY; here we just mean that the space has a K\"ahler metric with SU(3) holonomy, and in particular Ricci-flat. The conical requirement means that it can be written as $ds^2_{Y}= dr^2 + r^2 ds^2_{L_5}$ for a certain choice of coordinate $r$ and a five-manifold $L_5$, which we will call base (or link). $L_5$ is by definition a Sasaki--Einstein manifold. 

By the celebrated Yau's theorem, a compact K\"ahler manifold admits a Ricci-flat metric if and only if it has vanishing first Chern class; this is equivalent to the canonical bundle $K$, the bundle of $(3,0)$ forms, being trivial. In the non-compact case, however, such a simple criterion has been lacking. A singularity on which $K$ is trivial is called Gorenstein. In general, it is not true that any conical K\"ahler Gorenstein space admits a Ricci-flat metric: several obstructions to this were found in \cite{gauntlett-martelli-sparks-yau}. On the other hand, in the toric case there are no obstructions and the statement is true \cite{futaki-ono-wang}. (For a review of Sasaki--Einstein circa 2010, see \cite{sparks-rev}.) 

Recently, a criterion was proven \cite{collins-szekelyhidi}, called \textit{K-stability}, that guarantees the existence of a Calabi--Yau metric on a Gorenstein singularity, or equivalently of a Sasaki--Einstein metric on a five-manifold with positive curvature. This was inspired by the recent progress in the existence of K\"ahler--Einstein metrics \cite{chen-donaldson-sun}. The criterion is roughly speaking a generalization of volume minimization \cite{martelli-sparks-yau,martelli-sparks-yau2}, which identifies a conical Calabi--Yau metric among the set of conical complex ones and is the holographic dual of $a$-maximization \cite{intriligator-wecht}. While ordinary volume minimization requires varying among complex metrics on the same manifold, K-stability requires looking also at degenerations of the manifold. While in general one might not know how many such degenerations exist, in presence of two $\mathrm{U}(1)$ symmetries (rather than three as in the toric case) techniques exist \cite{altmann-hausen,ilten-suss} that insure that only a finite number of degenerations should be checked; we will review these techniques below. 

A tentative field-theoretic interpretation was proposed in \cite{collins-xie-yau} as a ``generalized $a$ maximization'' which instructs us to consider possible degenerations of the chiral ring. In fact the degenerations are associated to U(1) actions that are not symmetries except in certain limits; in a sense K-stability gives a very concrete realization to the idea of emergent IR symmetries.  

Thus, K-stability provides a way to produce new Sasaki--Einstein metrics. For example, \cite{collins-szekelyhidi} were able to find infinitely many new such metrics on $S^5$. It is now natural to want to develop the AdS$_5$/CFT$_4$ correspondence on these new metrics, and to find the corresponding quivers.

Finding the quiver corresponding to a singularity is not in general an easy task. In the toric case, an algorithm to do so was proposed in \cite{hanany-kennaway,franco-hanany-kennaway-vegh-wecht}, involving dimer models; recall that this was the case where a Calabi--Yau metric was already guaranteed to exist \cite{futaki-ono-wang}. In the non-toric case, which is now made accessible by K-stability, the dimer methods are not applicable. 

However, a different technique exists to find the quiver, called \textit{non-commutative crepant resolution (NCCR)} \cite{vdb-nccr}; importantly, it can also be applied to the non-toric case. It is based on algebraic ideas such as the one of \textit{matrix factorizations}, and is supposed to formalize the idea of trivial canonical bundle in terms of the path algebra of the quiver. Good introductions can be found in \cite{aspinwall-morrison-quivers,wemyss-lec}. For example, for a certain broad class of singularities, called ``compound $A_m$'' or $cA_m$ for short, it has already been reduced to an easy algorithm \cite{iyama-wemyss-cAn}, which produces quivers which further generalize the ``generalized conifolds'' of \cite{gubser-nekrasov-shatashvili}. In other cases computations are harder, but in principle still algorithmic.

Roughly speaking, these two separate developments can be viewed as progress on the complex and K\"ahler side of non-compact CYs. In this paper we put these two strands together to produce new AdS$_5$/CFT$_4$ duals. For simplicity we look at hypersurface singularities, namely singular spaces defined by a single equation $p(x,y,z,t)=0$ in $\mathbb{C}^4$. We look for examples where the K-stability test succeeds (and thus a Sasaki--Einstein metric is proven to exist) and an NCCR can be found. We exhibit a few new examples as a proof of concept. 

For instance among the $cA_m$ cases we find that the the singularity $uv+ z^p + t^p=0$ is both K-stable and has an NCCR. Outside the $cA_m$ class, where as we mentioned computations are more difficult, we find two examples of compound $D_4$ type, where the NCCR method reproduces quivers recently obtained in \cite{amariti-fazzi-mekareeya-nedelin} by abelian moduli space methods, and where we find that K-stability is also satisfied.

Rather strikingly, the NCCR and K-stability methods rarely agree with each other. If we start with a class where an NCCR exists, we find that it is rarely K-stable, and vice versa. It is natural to wonder what to do with the many singularities where only one of the criteria succeeds; we will come back on this point at the end of our investigation.

This paper is organized as follows. In section \ref{sec:k} we review K-stability and its application to SCFTs; in particular in subsections \ref{sub:fut}, \ref{sub:phys} and \ref{sub:torus} we review respectively its definition, its physics interpretation and the techniques that make it more manageable in presence of two $\mathrm{U}(1)$ symmetries. In section \ref{sec:nccr} we review the techniques of matrix factorizations and non-commutative crepant resolutions (NCCR), and introduce an algorithm due to Iyama and Wemyss to compute them in a class of singularities which from a physics point of view are similar to the generalized conifolds of \cite{gubser-nekrasov-shatashvili}; in section \ref{sec:cAm} we apply this algorithm to several types of K-stable singularities. In section \ref{sec:cD4} we start exploring the wider world of singularities where the Iyama--Wemyss algorithm does not apply, and reproduce some simple quivers with two or three nodes which have appeared very recently in \cite{amariti-fazzi-mekareeya-nedelin}. In section \ref{sec:conc} we will draw some conclusions from our investigations.


\section{Sasaki--Einstein manifolds and K-stability} 
\label{sec:k}

In this section, we will review the K-stability criterion for the existence of Sasaki--Einstein metrics \cite{collins-szekelyhidi}. First in sections \ref{sub:se} and \ref{sub:scft} we give a lightning review of well-known material about Sasaki--Einstein's and their dual SCFTs. In section \ref{sub:fut} we review K-stability, and in section \ref{sub:phys} we talk about its physics interpretation. In section \ref{sub:torus} we review techniques (mainly from \cite{altmann-hausen,ilten-suss}) to deal with manifolds which are non-toric but have only two $\mathrm{U}(1)$ actions, and which make K-stability more amenable to computations. Finally in section \ref{sub:k-ex} we review how K-stability applies to the class of so-called Brieskorn--Pham singularities.  

\subsection{Sasaki--Einstein threefolds} 
\label{sub:se}

We start with a lightning review of Sasaki--Einstein geometry for later reference. This material is well-known; for more details on Sasaki--Einstein geometry see \cite{sparks-rev}.

As we mentioned in the introduction, many AdS$_5$ solutions in IIB can be obtained by placing a stack of $N$ D3-branes at the tip of a conical CY threefold singularity $Y$. Recall that the CY condition can be formulated as the presence of a complex three-form $\Omega$ and two-form $J$, both non-degenerate and defining the same volume form ($\frac16 J^3 = -\frac i8 \Omega \wedge \bar \Omega = \mathrm{vol}_6$) and both closed ($dJ= d \Omega =0 $). The word ``conical'' means that the metric is of the form $dr^2+ r^2 ds^2_{L_5}$, where $L_5$ is a five-manifold called ``link''; one often writes $Y=C(L_5)$.  (Not all AdS$_5$ solutions are of this type, but in this paper we will restrict ourselves to this case.) Their back-reaction modifies the metric, and upon taking the near-horizon limit one obtains \cite{maldacena,morrison-plesser} a solution of the type AdS$_5 \times L_5$. In this paper we will focus on manifolds defined by a single polynomial equation 
\begin{equation}
	p(x,y,z,t)=0
\end{equation}
in $\mathbb{C}^4$. The holomorphic form is then given by the Poincar\'e residue expression:
\begin{equation}\label{eq:PR}
	\Omega= \frac{dx \wedge dy \wedge dz \wedge dt}{dp}= \frac{ dy \wedge dz \wedge dt}{\partial p/ \partial x}\,.
\end{equation}

By definition, $L_5$ is called Sasaki--Einstein if and only if $Y$ is Calabi--Yau. The holomorphic form $\Omega$ on $Y$ is the $(3,0)$-form of a complex structure $I$. The latter can be used to define a one-form and a vector field via
\begin{equation}
	\eta \equiv i (\overline{\partial} - \partial) \log r = I\left(\frac{dr}{r}\right) 
	\, ,\qquad
	\xi \equiv J(r \partial_r)\ .
\end{equation}
$\xi$ is called the Reeb vector. One can then reduce the forms on $Y$ to forms on $L_5$ via $\Omega= r^2(dr+ i r\eta)\wedge \omega$, $J= r dr \wedge \eta + r^2 j$; the two-forms $\omega$ and $j$ on $L_5$ then have to satisfy 
\begin{equation}
\begin{split}
	j \wedge \omega =0 \, & ,\qquad 2 j^2 = \omega \wedge \bar \omega \, ,\qquad \eta \wedge j^2= \mathrm{vol}_5\ ,\\
& d \eta = 2 j  \, ,\qquad d \omega = 3i \eta \wedge \omega \,.
\end{split}
\end{equation}
These relations can be taken as an alternative definition of a Sasaki--Einstein.

The orbits of the Reeb vector field $\xi$ can be compact or non-compact. If they are compact, $L_5$ is called semi-regular and is an $S^1$-fibration over a K\"ahler--Einstein $M_4$, possibly with orbifold singularities; if there are no orbifold singularities $L_5$ is called regular. This case is not very common, since there are very few K\"ahler--Einstein four-manifolds: $\mathbb{CP}^2$, in which case $L_5=S^5$ or $S^5/\mathbb{Z}_3$; $\mathbb{CP}^1\times \mathbb{CP}^1$, in which case $L_5= T^{1,1} \equiv \SU(2)/\U(1)\times \U(1)$ and $C(L_5)$ is called the conifold; and the del Pezzo surfaces dP$_k$ with $k\ge 3$. The conifold can also be described as the locus in $\mathbb{C}^4$ cut by the single quadric equation
\begin{equation}\label{eq:con-eq}
	x^2 + y^2 + z^2 + t^2 = 0 \,,
\end{equation}
which we will use as a running example. If the orbits are non-compact, $L_5$ is called irregular; this is by far the most common case, as first demonstrated in \cite{gauntlett-martelli-sparks-waldram-Ypq} with the discovery of the $Y^{p,q}$ metrics. Many more irregular examples can be produced by considering toric constructions and using the above-mentioned existence theorem \cite{futaki-ono-wang}. This case is well-understood and for this reason we will not consider it much in this paper. 

The isometry group of a compact manifold is compact; so when the orbits of $\xi$ are non-compact, there is in fact more than one Killing vector, of which $\xi$ is a linear combination with irrational coefficients. Other than in a few special cases, the isometry group is a torus
\begin{equation}\label{eq:torus}
	T\equiv \mathrm{U}(1)^r\,.
\end{equation}
 Again using the complex structure it follows that there is a $(\mathbb{C}^*)^r$ action on the manifold. One sometimes calls \textit{complexity} the (complex) dimension of the manifold minus $r$; in our case, $3-r$. For example, in the toric case $r=3$ and the complexity is zero. In what follows we will mostly deal with cases of $r=2$ and complexity one.

The Reeb vector $\xi$ also allows to compute easily the volume of $L_5$ via Duistermaat--Heckman localization \cite{martelli-sparks-yau,martelli-sparks-yau2,bergman-herzog}.
Another way of computing the volume is via the Hilbert series (HS) of the threefold, $H(u)\equiv \dim (H_d) u^d$, where $\dim (H_d)$ is the dimension of the space of holomorphic functions of degree $d$ under $\xi$, namely functions $h$ such that $L_\xi h= d h$. It turns out that
\begin{equation}\label{eq:Hs}
	H(e^{-s}) =  \frac{2 a_0}{s^3} + \frac {a_1}{s^2} + O(s^{-1})  \, ,\qquad a_1=3 a_0\,,
\end{equation}
in conventions where (\ref{eq:PR}) has degree $d=3$. The volume is then given by
\begin{equation}\label{eq:vol-a0}
	\mathrm{Vol}(L_5)= 2 a_0\mathrm{Vol}(S^5)=2a_0 \pi^3\,.
\end{equation}
For a hypersurface $p=0$ in $\mathbb{C}^4$, the Hilbert series can be computed in terms of the degrees of the coordinates 
\begin{equation}\label{eq:zi}
	z_i=(x,y,z,t)
\end{equation}
of $\mathbb{C}^4$. If we define $w_i>0$ the (positive) degrees of $z_i$ under $\zeta$ and  $w_p$ the total degree of $p$ (which is homogeneous under $\zeta$), we have
\begin{equation}\label{eq:Hs-hyp}
	H(e^{-s})=\frac{1-e^{-w_p s} }{\prod_i (1-e^{w_i s})}\,.
\end{equation} 
From this we can expand in $s$ and find
\begin{equation}
	a_0 = \frac{w_p}{2\prod_i w_i}\,.
\end{equation}
Rather than using the $w_i$ directly, it is sometimes simpler to introduce a vector $\zeta= b \xi$ proportional to the Reeb vector, and determine $b$ by fixing $\Omega$ to have degree 3. Let us call $\alpha_i$ and $\alpha_p$ the degrees of $z_i$ and $p$ respectively under $\zeta$; then $w_i = b \alpha_i$, $w_p = b \alpha_p$. The condition that $\Omega$ should have degree 3 (or that $a_1=3a_0$ in (\ref{eq:Hs})) fixes $b=\frac3{-\alpha_p+ \sum_i \alpha_i}$, and
\begin{equation}\label{eq:a0-alpha}
	 \qquad a_0 = \frac{\alpha_p}{2b^3\prod_i \alpha_i}= \frac{\alpha_p}{54\prod_i \alpha_i}\left(-\alpha_p + \sum_i \alpha_i\right)^3\,.
\end{equation}
The advantage of this point of view is that we do not have to worry about normalizing the $\alpha_i$, $\alpha_p$. 

For example, for the conifold (\ref{eq:con-eq}) we have $\alpha_i=(2,2,2,2)$ and $\alpha_p=4$; from (\ref{eq:a0-alpha}) and (\ref{eq:vol-a0}) we get $a_0= \frac{8}{27}$ and $\mathrm{Vol}(T^{1,1})=\frac{16}{27}\pi^3$.

All this also gives a way to compute the Reeb vector $\xi$. Considering a general linear combination $\sum_i \ell_i K_i $ of the generators of the isometry torus $\mathrm{U}(1)^r$, the volume $\mathrm{Vol}(L_5)$ will depend on the coefficients $\ell_i$; the Reeb vector $\xi$ is then found by minimizing the volume with respect to the $\ell_i$ \cite{martelli-sparks-yau2}. 


\subsection{Superconformal models} 
\label{sub:scft}

Let us also review briefly some aspects of superconformal theories (SCFTs) in four dimensions.

The class of theories we consider in this paper are ${\mathcal N}=1$ quiver theories: namely, they have several vector multiplets with gauge groups $\mathrm{SU}(N_i)$, and chiral multiplets transforming in various bifundamental and adjoint representations. (We will not consider matter in the fundamental representation of a gauge group.) A necessary condition for the theory to be superconformal is that the beta functions of all the gauge groups vanish:
\begin{equation}\label{eq:beta}
\beta_i \equiv N_i + \sum_{e_i} N_i (R_{e_i}-1) + \frac{1}{2}\sum_{a:i\to j} N_j (R_{a}-1) = 0 \ 
\end{equation}
where $e_i$ are the adjoint chirals, $a:i\to j$ denote the bifundamentals, and $R_{e_i}$, $R_a$ are their charges under the $\mathrm{U}(1)$ R-symmetry. Depending on the model, one can then sometimes argue that a choice of R-charges exists such that (\ref{eq:beta}) can be satisfied. One typically treats the $R_{e_i}$ and $R_a$ as functions of the gauge couplings and superpotential coefficients; a counting argument then tells us if a solution is expected to exist. A more rigorous and laborious way of proceeding that is often used in the literature (see e.g. \cite{leigh-strassler}) is to proceed in steps, starting from a model where there is no superpotential and introducing terms in the superpotential step by step. At each step one first $a$-maximizes among the possible non-anomalous R-charge assignments, and then checks which operators are relevant with the $a$-maximized values of the R-charges; switching these operators on makes one flow to a more complicated model. In a sense the K-stability procedure in this paper is a formalization of these ideas.

In any case, it is not easy to show that a SCFT exists in a completely rigorous fashion purely from field theory arguments. So it is helpful when a model has a holographic dual, for which the existence of a Sasaki--Einstein metric on the dual geometry can be proven. We will use this perspective in this paper.

Once a SCFT exists, anomalies give an interesting measure of the number of degrees of freedom it contains. For four-dimensional conformal theories, there are two possible Weyl anomalies, called $a$ and $c$. With $\None$ supersymmetry, they can be expressed in terms of R-symmetry anomalies \cite{anselmi-freedman-grisaru-johansen}:
\begin{equation}\label{eq:4d-ac}
a=\frac{3}{32}(3 \Tr R^3 - \Tr R)\ , \quad \quad c =a-\frac{1}{16}\Tr R = \frac{3}{32}\left(3 \Tr R^3 - \frac{5}{3}\Tr R\right)\ ,
\end{equation}
where $R$ is the R-symmetry generator. For a quiver theory with $\SU(N_i)$ gauge groups (and no fundamentals), this gives 
\begin{subequations}\label{eq:4dac}
\begin{align}
a &=\frac{3}{32}\left(2 \sum_{i} (N_i^2-1) + \sum_{a:i \to j} N_i N_j \left[ 3 (R_a-1)^3 - (R_a-1) \right] \right. + \nonumber \\ &\ \ \ \, +\left. \sum_{e_i} (N_i^2-1) \left[ 3 (R_{e_i}-1)^3 - (R_{e_i}-1) \right]  \right)\ , \label{eq:a-4dquiv}\\ 
c&=\frac{3}{32}\left(\frac{4}{3} \sum_{i} (N_i^2-1) + \sum_{a:i \to j} N_i N_j \left[ 3 (R_a-1)^3 - \frac{5}{3}(R_a-1) \right] \right. + \nonumber \\ &\ \ \ \, +\left. \sum_{e_i} (N_i^2-1) \left[ 3 (R_{e_i}-1)^3 - \frac{5}{3}(R_{e_i}-1) \right]  \right)\ .
\label{eq:c-4dquiv}
\end{align}
\end{subequations}
For SCFTs with a weakly-coupled gravity dual, $a$ and $c$ should be equal at large $N$; interestingly, for a quiver theory (without fundamental matter) this follows from \eqref{eq:4dac} \cite{benvenuti-hanany}.

The $a$ anomaly is related to the volume of the gravity dual $L_5$ \cite{gubser-vol,henningson-skenderis}: 
\begin{equation}\label{eq:acharge-vol}
	a= \frac{\text{Vol}(S^5)}{\text{Vol}(L_5)}\, a_{{\mathcal N}=4\, \text{SYM}}\,.
\end{equation}
(Common conventions give $a_{{\mathcal N}=4\, \text{SYM}}=\frac{N^2}4$ for $\mathcal{N}=4$ SYM with $\SU(N)$ gauge group.)
The volume minimization of $L_5$ is then dual to the statement that the R-charge assignment should maximize $a$ \cite{intriligator-wecht}.


\subsection{K-stability and the Futaki invariant} 
\label{sub:fut}

We now describe the K-stability procedure; for a more thorough introduction, see for example \cite{collins-phd}.

The idea of stability has a long history. System of PDEs can often be separated into holomorphic and real equations; in supersymmetric theories these can sometimes be interpreted as F- and D-term equations respectively. Holomorphic equations can be solved easily with algebraic-geometrical methods; for the real equations, one can sometimes use the action of the complexification $G_\mathbb{C}$ of a symmetry group to try to reach a solution. Some orbits contain such a solution and are called ``stable'', while others do not, the $G_\mathbb{C}$ action degenerating to other, simpler orbits. A notable example is the self-duality equation $F=*F$ in four Euclidean dimensions, which can be separated into a holomorphic part $F_{2,0}=0$ and a real part $J\cdot F_{1,1}=0$; the latter can be proven to be solved when a certain stability test succeeds, involving sub-bundles of the bundle of which $F$ is a curvature. The general story here leads to the Donaldson--Uhlenbeck--Yau equations \cite{donaldson,uhlenbeck-yau}. A similar story was later conjectured for the existence of K\"ahler--Einstein metrics \cite{yau-ke} and more recently proven in the existence direction in \cite{chen-donaldson-sun} (with an earlier necessity result for example in \cite{stoppa}). The idea is that this time the complexified symmetry action makes the manifold itself degenerate to another, simpler manifold. Here we will need a variant which applies to Sasaki--Einstein metrics \cite{collins-szekelyhidi}.

The degenerations we will need are called \textit{test configurations}. As we anticipated, such a degeneration is usually obtained by an action on the coordinates, which will be a $\mathbb{C}^*$ action.  In our cases, where we have a torus $T$ of symmetries \eqref{eq:torus}, we only need to take into account \textit{$T$-equivariant} test configurations, namely those that are generated by actions that commute with $T$. (From now on we will drop the qualifier ``$T$-equivariant'' and simply call this a ``test configuration''.) Such an action $\mathbb{C}_\lambda$ might then have a degeneration in the limit where the generator $\lambda$ goes to zero.

For example, for the conifold (\ref{eq:con-eq}), one such action might be 
\begin{equation}\label{eq:deg}
	(x,y,z,t)\to \lambda\cdot(x,y,z,t)\equiv (x,y,z,\lambda t)\,.
\end{equation}
For any $\lambda $ this takes us to a new equation; for $\lambda\neq0$ this is isomorphic to the original conifold, but for $\lambda=0$ we have the degeneration
\begin{equation}\label{eq:deg-con}
	x^2+y^2+z^2=0\,.
\end{equation}
The action (\ref{eq:deg}) has now become a symmetry on the degeneration (\ref{eq:deg-con}).

The formal definition is as follows. A $T$-equivariant test configuration of $Y$ is an embedding $Y \hookrightarrow \mathbb{C}^D$ on which $T$ acts as a unitary representation, together with a one-parameter subgroup $C_\lambda : \cc^* \to \U(D)^T$, namely one which is unitary and commutes with $T$.  This action takes $Y$ to a $Y_\lambda$; for $\lambda\neq0$, these are all isomorphic to each other. On the other hand, $Y_0$ is special: it is left invariant by $C_\lambda$, and can be different from $Y$. This $Y_0$ is called the ``central fiber'' of the test configuration. (The name comes from thinking of the $Y_\lambda$ as fibers of a flat fibration over a copy of $\mathbb{C}$ parameterized by $\lambda$.) We also require that $Y_0$ be normal: namely, its ring $R_0$ of functions is an integrally closed domain, or in other words there are no solutions $f$ to an algebraic equation $f^n + \lambda_{n-1} f^{n-1}+ \ldots f_0=0$ where $\lambda_i \in R_0$ and $f$ is not in $R_0$ but rather a rational function. A classic example of non-normal variety is the cusp
\begin{equation}\label{eq:cusp}
	x^2-y^3=0\,.
\end{equation}
Indeed the rational function $f=\frac {y^2}x$ satisfies $f^2-y=0$: $\left(y^2/x\right)^2-y=\frac y{x^2}(y^3-x^2)$. On the other hand, the conifold $x^2+y^2+z^2+t^2=0$ is normal.

Since $Y_0$ is left invariant by $\lambda$, on it we have one more U(1) than on the original $Y$. It is then natural to wonder whether this extra U(1) changes volume minimization. This ``generalized volume minimization'' is then the idea of K-stability. Since the Reeb vector already minimizes the volume in the space of all the generic U(1) symmetries present at generic $\lambda$, it is enough to minimize with respect to variations that include this new symmetry $\lambda$ as well. (We are calling $\lambda$ both the $\mathbb{C}^*$ action and the generator of the U(1) inside it, hoping that this will not generate confusion.) One sees a parallel \cite{collins-xie-yau} with the field theory idea that an emergent symmetry in the IR might invalidate $a$-maximization computations; we will expand on this comment in section \ref{sub:phys}. 

Concretely, one performs this generalized minimization by computing the \textit{Futaki invariant}
\begin{equation}\label{eq:Fut}
	\mathrm{Fut}(\xi,\lambda) \equiv \left(\partial_\epsilon a_0 + a_0 \partial_\epsilon\left(\frac{a_1}{a_0}\right)\right)_{\epsilon=0}\,,
\end{equation}
where $a_i=a_i(\xi + \epsilon \lambda)$. If $\mathrm{Fut}(\xi,\lambda) \le 0$, then one says that the test configuration induced by $\lambda$ destabilizes $Y$. This is an obstruction to the existence of a conical Calabi--Yau metric on it. On the other hand, the converse is also true \cite[Thm 1.1]{collins-szekelyhidi}: if $Y$ is not destabilized by any test configuration, then $Y$ is said to be \textit{K-stable} and there exists a conical Calabi--Yau metric on it; in other words, there is a Sasaki--Einstein metric on $L_5$.

To see the connection with generalized volume minimization, notice from (\ref{eq:Hs}) that $a_1=3a_0$; then
\begin{equation}\label{eq:int-Fut}
	\frac{\mathrm{Fut}}{a_0} = \partial_\epsilon \left(\log a_0 + 3\log\left( \frac{a_1}{a_0}\right)\right)_{\epsilon=0}= \partial_\epsilon\log\left(a_0\left( \frac{a_1}{a_0}\right)^3 \right)_{\epsilon=0}\,.
\end{equation}
We can view the $(a_1/a_0)^3$ factor as the effect of renormalizing the degrees so that $a_1/a_0$ remains equal to 3 while varying, in a similar logic to the $b^{-3}$ factor in (\ref{eq:a0-alpha}) in our hypersurface case. If we denote by $t_i$ the degrees of the action of $\lambda$,
\begin{equation}\label{eq:ti}
	\lambda\cdot z_i = z_i^{t_i}
\end{equation}
(in the ordering (\ref{eq:zi})), in (\ref{eq:int-Fut}) we are computing the derivative of the logarithm of
\begin{equation}
	a_0 \left( \frac{a_1}{a_0}\right)^3= \frac{w_p+ \epsilon t_p}{2 \prod_i (w_i + \epsilon t_i)}\left(\sum_i(w_i+ \epsilon t_i)- (w_p + \epsilon t_p)\right)^3\,.
\end{equation}
An equivalent point of view, promoted in \cite{collins-szekelyhidi,collins-xie-yau}, is that one varies by rescaling $\xi$ at the same time as adding the new generator $\lambda$:
\begin{equation}
	\mathrm{Fut}(\xi,\lambda)\equiv \left(\frac{\partial}{\partial \epsilon}a_0(\xi + \epsilon (\lambda-\alpha \xi))\right)_{\epsilon=0} \, ,\qquad \left(\frac{\partial}{\partial \epsilon} \frac{a_1}{a_0}\right)_{\epsilon=0}=0\,,
\end{equation}
where $\alpha$ is fixed by the second equation.

From any of these points of view, after some manipulations we obtain 
\begin{equation}\label{eq:exp-Fut}
	\frac{\mathrm{Fut}}{a_0}= - t_p + \sum_i t_i +\frac13\left(\sum_i \alpha_i - \alpha_p\right)\left(\frac{t_p}{\alpha_p}-\sum_i \frac{t_i}{\alpha_i}\right)
\end{equation}
for the hypersurface case of interest in this paper.

Notice that the degeneration $Y_0$ has a chance of being a Calabi--Yau itself. By construction $Y_0$ has one more $\mathbb{C}^*$ action than the original $Y$. Given that we need to have at least one $\mathbb{C}^*$ action, we only have three cases, each of which can degenerate to the next: 
\begin{equation}\label{eq:hierarchy}
		\begin{array}{c}
			\text{one } \mathbb{C}^* \text{ action}\\ \text{(complexity two)}			
		\end{array}\ 
		\longrightarrow \
		\begin{array}{c}
			\text{two } \mathbb{C}^* \text{ actions}\\ \text{(complexity one)}			
		\end{array}
		\ 
		\longrightarrow \
		\begin{array}{c}
			\text{three } \mathbb{C}^* \text{ actions}\\ \text{(toric)}
		\end{array}\ .
\end{equation}
In particular, if $Y$ is toric, it has no possible degenerations, since it has already the maximum number of $\mathbb{C}^*$ actions in three dimensions. Indeed in the toric case the existence of a Sasaki--Einstein metric is guaranteed \cite{futaki-ono-wang}. On the other hand, if $Y$ has complexity one, the degeneration $Y_0$ is toric. 
 
Checking positivity of the Futaki invariant for all possible test configurations might seem like a daunting task. Fortunately, we will see in section \ref{sub:torus} that for complexity one only a finite number of configurations has to be checked. This is the case we will restrict in most of this paper. The complexity-two case is more complicated, although it might be amenable to similar methods in the future.


\subsection{Physical interpretation} 
\label{sub:phys}

While K-stability comes from geometry, it is natural to try and translate the idea into physics. 

We have seen that K-stability requires us to look at $\mathbb{C}^*$ actions which are not symmetries of the original $Y$, to consider the new threefolds $Y_0$ one obtains by letting this action degenerate, and check that the Futaki invariant is positive. We have also seen that this can be interpreted as checking volume minimization on $Y_0$, which has more $\mathrm{U}(1)$ symmetries than $Y$. 

The holographic dual of volume minimization is $a$-maximization \cite{martelli-sparks-yau}, which says that the choice of R-symmetries among the $\mathrm{U}(1)$ actions should maximize $a$ \cite{intriligator-wecht}. One then wants to interpret K-stability as a ``generalized $a$-maximization'' \cite{collins-xie-yau} which requires one to check that $a$ is maximized even taking into account $\mathrm{U}(1)$ actions which are not symmetries. It is natural to also conjecture \cite{collins-xie-yau} that K-stability holds directly for the chiral ring of any putative SCFT, even without a holographic dual. This would mean that a theory with chiral ring $R$ is an SCFT if and only if all its degenerations $R_0$ (generated by additional $\mathrm{U}(1)$ actions as in section \ref{sub:fut}) do not have a higher $a$. (A related conjecture appeared in \cite{benvenuti-giacomelli}. We will come back to it in our conclusions.)

More precisely, when the Futaki invariant is positive, it signals that one can make $a_0$ smaller (and hence the $a$ anomaly larger) by varying with respect to the additional $\mathrm{U}(1)$ associated to the test configuration, by making $\epsilon$ positive. On the other hand, when the Futaki is negative, one cannot do that and there is no reason to think that $a$ can be made larger even by including the extra $\mathrm{U}(1)$. Notice that we \emph{cannot} vary in the direction of negative $\epsilon$: it would correspond to a choice of R-charges which contradicts the original assumption, i.e. that the chiral ring degenerates to $R_0$.

The reason one wants to maximize with respect to the additional $\mathrm{U}(1)$ present in the degenerate chiral ring $R_0$ is a manifestation of a well-known caveat about $a$-maximization: namely, that one should take into account the possibility of emerging symmetries in the IR. When generalized $a$-maximization fails, some terms of the superpotential $W$ have gone to zero in the IR, and an extra $\mathrm{U}(1)$ emerged. The theory with chiral ring $R$ is not itself an SCFT: it flows in the IR to the theory with simpler $W$ and with chiral ring $R_0$, which has an additional $\mathrm{U}(1)$ symmetry. This second theory might in fact also not be an SCFT, but degenerate in turn to another SCFT, in a field-theory counterpart of the hierarchy (\ref{eq:hierarchy}). 

Our general discussion so far might have given the impression that generalized $a$-maximization for $Y$ is just the same as ordinary $a$-maximization on the central fiber $Y_0$. However, a crucial difference is that with generalized $a$-maximization we can only vary R-charges compatibly with the assumption that the $\lambda$ action makes the chiral ring degenerate to that of $Y_0$; this effectively creates a boundary in the allowed minimization space. This is related to our observation above, that when the Futaki invariant is negative one cannot go in the direction of negative $\epsilon$ to lower $a_0$ (and raise $a$). We will see this in more detail in the examples of section \ref{sub:uv-laufer} and \ref{sub:laufer}.

As we mentioned at the end of section \ref{sub:fut}, in our paper we will mostly focus on theories with two $\mathrm{U}(1)$'s; these are either SCFTs themselves, or flow in the IR to toric SCFTs. Occasionally we will also speculate on examples with only one U(1), most notably in section \ref{sub:laufer}.


\subsection{Torus actions with complexity one} 
\label{sub:torus}

In the toric case (i.e.~when there is a $T=\mathrm{U}(1)^3$ of isometries) the geometry of a threefold can be summarized very effectively by the so-called toric diagrams and by toric polytopes. These are two ways, dual to each other, to represent visually the $T$ action in the various coordinate patches.

When there are fewer isometries, these methods are still partially available. We will focus here on the case with complexity one, i.e.~when $T=\mathrm{U}(1)^2$. (We will focus on the threefold case, but these techniques can be applied to any dimension.) This topic has a long history; the reader may for example consult \cite{altmann-hausen}.\footnote{We would like to thank N.~Ilten, G.~Sz\'ekelyhidi and especially H.~S\"u\ss~for illuminating email correspondence about several aspects of this topic.}

Our manifold $Y$ can be realized as a fibration over a Riemann surface $B$, with the $T=\mathrm{U}(1)^2$ acting on the fiber $F$, and some special points $p_i\in B$ where $F$ changes. The data of the $T$ action are summarized by the $p_i$ and some polytopes $\Delta_i$, of dimension 2. One sometimes also introduces the formal sum $\sum_i p_i \Delta_i$, called \textit{proper polyhedral (pp) divisor}. 

The methods in \cite{ilten-suss} can be used to compute combinatorially all test configurations in terms of the $p_i$ and $\Delta_i$. In this paper however we will only use this result to count the number of test configurations, and then find them explicitly by hand as in \cite{collins-szekelyhidi}. For this, one has to compute certain linear piecewise functions $\Psi_i$, and perform a certain test which we will introduce.

Let us explain these methods concretely using an example: the threefold defined by the equation
\begin{equation}\label{eq:p-uv-laufer}
	p=x^2 + y^3 + z^2 t=0\,.
\end{equation}
(We will analyze the dual theory in section \ref{sub:uv-laufer}.) 

First let us try to find some test configurations by hand. One obvious idea is to make disappear one of the three monomials in (\ref{eq:p-uv-laufer}). For example, in the notation of (\ref{eq:ti}), the action $(0,0,0,1)$ leads to $x^2 + y^3 + \lambda^2 z^2 t$, whose central fiber $x^2+y^3$ is the cusp (\ref{eq:cusp}) and hence not normal, as remarked there. If we try with $(0,1,0,0)$, the central fiber is the ``Whitney umbrella''
\begin{equation}\label{eq:whitney}
	x^2+z^2t=0
\end{equation}
which is not normal because $f=\frac xz$ satisfies $f^2+t=0$. Finally we can try with $(1,0,0,0)$, which leads to the central fiber $y^3+z^2t=0$; also this is not normal, since $f=\frac{y^2}z$ satisfies $f^2-yt=0$. So the naive attempts at getting a valid test configuration fail because of non-normality of the central fiber. One could imagine more elaborate actions, for example non-diagonal ones. The point of the methods we will explain now is precisely that it gives a systematic way of finding all test configurations, without having to guess.

\subsubsection{Fibration and special points} 
\label{ssub:pi}

The first step is to find the two $\mathbb{C}^*$ symmetries of this equation: they are given by the charge matrix
\begin{equation}\label{eq:C*reeb-UV}
F=\left(
\begin{array}{cccc}
3 & 2 & 3& 0 \\
0 & 0 & -1& 2 
\end{array}
\right)\,.
\end{equation}  
which represents the action respectively on $(x,y,z,t)$.
As usual in toric geometry, we need to compute its kernel, namely a matrix $P$ such that $F \cdot P^\text{t}=0$; additionally, we will need a matrix $s$ such that $F \cdot s^\text{t} = 1$: 
\begin{equation}
	P = \left( 
	\begin{array}{cccc}
	-2 & 0 & 2 & 1\\
	-2 & 3 & 0 & 0 	
	\end{array}	
	\right)
	\, ,\qquad
	s= \left(\begin{array}{cccc}
		1& -1 & 0  & 0 \\
		1& 0  & -1 & 0
	\end{array}\right)\,.
\end{equation}

\begin{figure}[!ht]
\centering
\includegraphics[scale=.8]{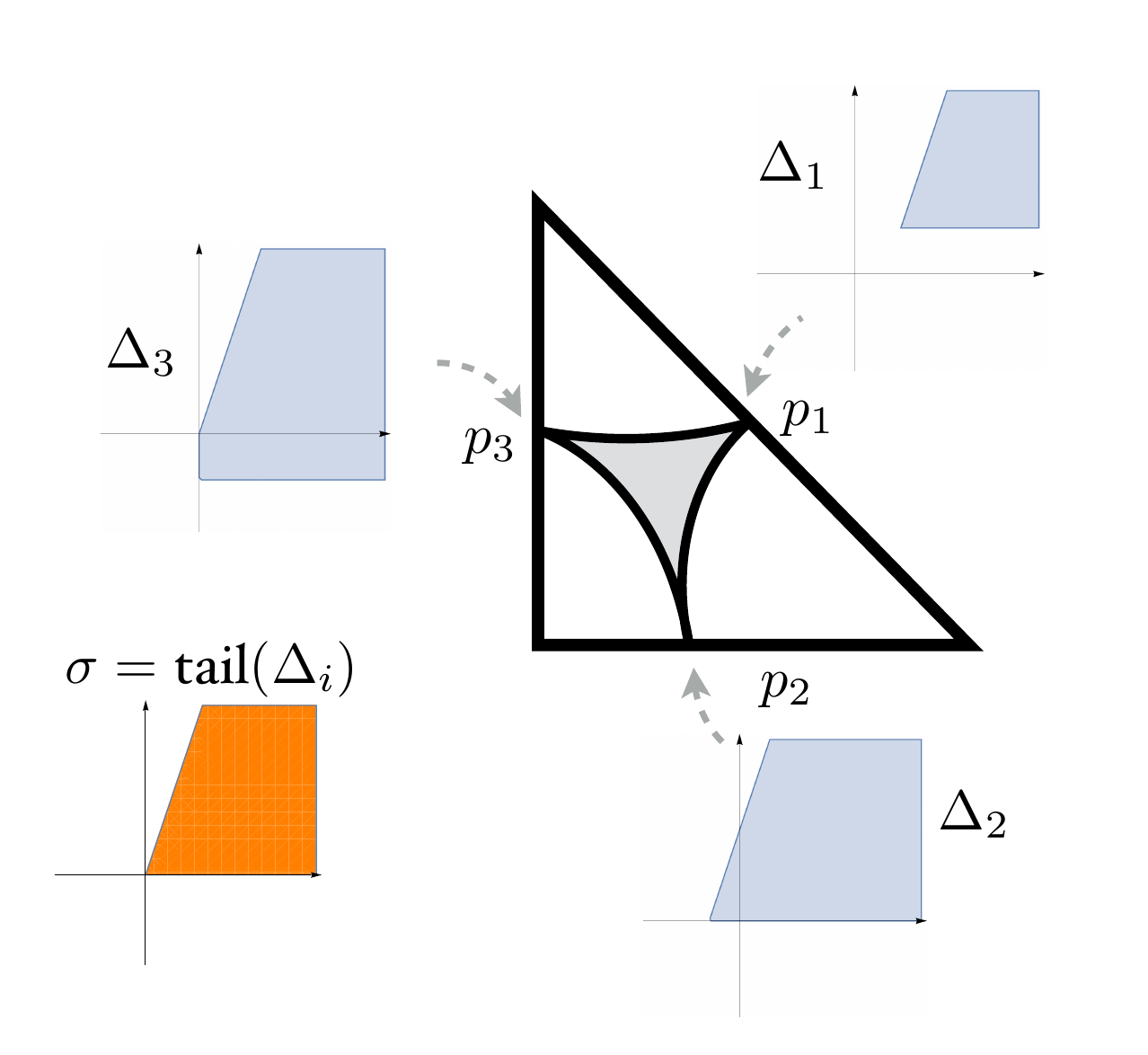}
\caption{The base $B$ and its associated polytopes for our example (\ref{eq:p-uv-laufer}).}
\label{fig:pp}
\end{figure}

To determine the base $B$, we view $P$ as giving column vectors in $\mathbb{C}^2$; the rays traced by these vectors give the fan of a toric manifold. In this case we see that the generators of these rays are the vectors $v_1={-1\choose -1}$, $v_2={0 \choose 1}$ and $v_3={1\choose 0}$ (repeated twice). We recognize the fan of $\mathbb{CP}^2$; the equation (\ref{eq:p-uv-laufer}) now gives a hypersurface inside it. To read it off, we map $\mathbb{C}^4$ to the affine coordinates in a chart of $\mathbb{CP}^2$ by using the rows of $P$: 
\begin{equation}
	(x,y,z,t)\mapsto \left(X\equiv \frac{y^3}{x^2}, Y \equiv \frac{z^2t}{x^2} \right)\,.
\end{equation} 
Then (\ref{eq:p-uv-laufer}) becomes in this chart the linear equation $1+X+Y=0$. It can be useful to projectivize this by introducing further homogeneous coordinates $(w_0,w_1,w_2)$ in $\mathbb{CP}^2$ such that $X=\frac{w_1}{w_0}$, $Y=\frac{w_2}{w_0}$; then the equation becomes $w_0+w_1+w_2=0$. Being of degree one, this cuts a Riemann surface of genus zero, which is the base $B$ we anticipated. By \cite[Cor.~5.8]{liendo-suss}, $B$ will in fact always have genus zero for the cases of interest in this paper, namely when the threefold $Y$ is a Calabi--Yau; more generally it might have higher genus if we are interested in Sasaki--Einstein manifolds of non-positive curvature. We show the situation schematically in figure \ref{fig:pp}, with the toric polytope of $\mathbb{CP}^2$ and $B$ depicted inside it.\footnote{This schematic depiction of $B$ simply tries to convey that it is topologically an $S^2$, and that it intersects each $D_i$ once. One can also think of the gray region as a so-called \textit{am\oe ba}, namely the image of $B$ under the toric fibration map over the triangle, whose generic fibers are $T^2$s.}

Let us now call $p_a$ the intersections of $B$ with the three toric divisors $D_i$ of $\mathbb{CP}^2$, namely the loci $\{w_i=0\}$, which are associated to the vectors $v_i$ above. In our example, since the equation is linear, each intersection consists of a single point; the resulting $p_1$, $p_2$, $p_3$ are shown in figure \ref{fig:pp}. More generally, one can have several intersections even when $B$ has genus zero. This can happen for example if it is inside a weighted projective space; in fact we will see such a case in section \ref{sub:gen}.  

We can now already explain the geometry of the degeneration. $Y$ already has two abelian isometries; the degeneration of a test configuration should have three, and thus be toric. In terms of the fibration we just described, $F$ is already toric; so the extra abelian isometry should somehow arise from a degeneration of the base $B$.  To see how this can happen, focus on how the test configuration's $\mathbb{C}^*$ action $\lambda$ acts on $B$. In the $\lambda\to 0$ limit, somehow $B$ has to acquire an isometry; in other words it has to become toric itself, so $B\to \mathbb{CP}^1$. Moreover, the $p_i$ will have to be collected at one of the two poles, which are the fixed points of the new, emerging abelian isometry. 
The positions of the $p_i$ on $B$ rescale under $\lambda$; in the degeneration limit $\lambda\to 0$, they will all go to the $z=0$ point in $\mathbb{CP}^1$, except if one or more points happened to be at $z=\infty$. So, from the point of view of $B$, the possible degenerations correspond to the possible choices of one of the $p_i$ we call $z=\infty$;\footnote{In principle one could also choose $z=\infty$ not to coincide with any of the $p_i$, but this never leads to a destabilizing degeneration; we thank H.~S\"u\ss~for explaining this to us.} all the other $p_i$ will coincide in the $\lambda\to 0$. 

In our example, we can let $z=\infty$ be $p_1$, and then in the $\lambda\to 0$ limit $p_2$ and $p_3$ will coincide; this will be a possible degeneration. There are two more possibilities, and we then see a possible total of three degenerations. 

As we mentioned, each degeneration $Y_0$ will be toric.  \cite{ilten-suss} gives a quick combinatorial way to compute the toric diagram of $Y_0$, and in particular a way to check that it is normal. In what follows we will only review the latter, which is enough for our purposes.


\subsubsection{Polytopes} 
\label{ssub:poly}

The combinatorial method of \cite{ilten-suss} requires the introduction of certain polytopes $\Delta_i$, associated to the $p_i$. The procedure is as follows: 
\begin{itemize}
	\item each $p_a$ is the intersection of $B$ with a toric divisor $D_i$; 
	\item take the counterimage of the corresponding fan vector $v_i$ under $P$, $P^{-1}(v_i)$, and intersect it with the positive quadrant $\mathbb{R}^4_{\ge 0}$;
	\item compute the image of $s$ on the resulting $\mathbb{R}^4_{\ge 0}\cap P^{-1}(v_i)$.                                     
\end{itemize}
Let us see this in our example. For $v_1={-1\choose -1}$:
\begin{subequations}\label{eq:Deltai}
\begin{equation}
\begin{split}
	&P\left(\begin{array}{c}
	  \alpha \\ \beta \\ \gamma \\ \delta   
	\end{array}\right)=
	{-2 \alpha + 2 \gamma + \gamma \choose -2 \alpha + 3 \beta }
	={-1 \choose -1}\
 \Rightarrow \ 
	P^{-1}(v_1)=
	\left\{ \left( \begin{array}{c}
		(1+2 \gamma + \delta)/2 \\ (2 \gamma + \delta)/3 \\ \gamma \\ \delta 
	\end{array}\right)\right\}\,;\\
	 &\mathbb{R}^4_{\ge 0}\cap P^{-1}(v_1)=   \left\{ \left( \begin{array}{c}
		(1+2 \gamma + \delta)/2 \\ (2 \gamma + \delta)/3 \\ \gamma \\ \delta 
	\end{array}\right)\,,\ 
	\begin{array}{c}
		   \gamma \ge 0 \\ \delta \ge 0
	\end{array}\right\} \, ;\\
	&\Delta_1 = s\left(\mathbb{R}^4_{\ge 0}\cap P^{-1}(v_1)\right)   = \left\{\left( \begin{array}{c}
		(3+2 \gamma + \delta)/6 \\ (1+\delta)/2 
	\end{array}\right)\,, \ 	\begin{array}{c}
			   \gamma \ge 0 \\ \delta \ge 0
		\end{array}\right\}= 
		{1/2 \choose 1/2}+ \sigma \,.
\end{split}	
\end{equation}
In the last step we have written $P_1$ as a ``Minkowski sum''\footnote{The Minkowski sum of two polytopes $P_1$, $P_2$ is defined as the set of vectors that can be written as $v_1 +v_2$ for some $v_i \in \Delta_i$. In this case $P_1$ is a single vector, and $P_2= \sigma$; the result is just a translation of $\sigma$.} of the single vector ${1/2 \choose 1/2}$ with the cone
\begin{equation}\label{eq:tailcone}
	\sigma \equiv \left\{ \tilde\gamma {1 \choose 0} + \tilde \delta {1\choose 3}\,,\ \tilde \gamma \ge 0
	\,,\ \tilde \delta \ge 0 
 	\right\}\,.
\end{equation}
The other two polytopes are computed in the same way, and are 
\begin{equation}
	\Delta_2 =   
	{-1/3\choose 0} + \sigma \, ,\qquad
	\Delta_3 =  
	\left\{{\gamma\choose 0}, \gamma\in[-1/2,0]\right\}  + \sigma\,.
\end{equation}
\end{subequations}
We show the $\Delta_i$ and $\sigma$ in figure \ref{fig:pp}.

We see that all the $\Delta_i$ can be written as Minkowski sums with the same cone $\sigma$, which is also called their \textit{tailcone} $\mathrm{tail}(\Delta_i)$. For any polytope $\Delta$, we can also define its tailcone as the cone of unbounded directions in $\Delta$: formally, $\mathrm{tail}(\Delta)=\{v | v'+ t v \in \Delta \forall v' \in \Delta, t \in \mathbb{R}_{\ge 0}\}$. We can also get $\sigma$ by intersecting the image of $F$ with the positive quadrant $\mathbb{R}^4_{\ge 0}$; in our case 
\begin{equation}
	\mathbb{R}^4_{\ge 0} \cap F(\mathbb{R}^2) = \left\{ \left(\begin{array}{c}
		3a \\ 2a \\ 3a-b \\ 2b
	\end{array}\right)\,, 3a\ge b\ge 0\right\}\,,
\end{equation}
which coincides with $\sigma$ in (\ref{eq:tailcone}).

We now introduce some piecewise-linear functions $\Psi_i$. These are defined on the dual $\sigma^\vee$ of $\sigma$, namely the set of vectors that have positive inner product with all vectors of $\sigma$, and are given by
\begin{equation}\label{eq:Psii}
	 \Psi_i \equiv \min_{v \in \Delta_i}(u \cdot v)\,.
\end{equation} 
A $\Psi_i$ is said to have \textit{integer slopes} if 
\begin{equation}\label{eq:int-slope}
	\forall \,v \in \sigma^\vee \ \text{with integer coefficients} \ \Psi_i(v) \in \mathbb{Z}\,.
\end{equation}

In our running example, $\sigma^\vee = \{(s,t)| s \ge 0, s+3t\ge 0\}$, and 
\begin{equation}\label{eq:Psii-ex}
	\Psi_1 = \frac12(s+t) \, ,\qquad \Psi_2 = -\frac s3 \, ,\qquad \Psi_3 = \left\{ \begin{array}{cc}
		-t/2 & \ \ t> 0\,, \\
		0 & \ \ t\le 0\,.
	\end{array}\right.
\end{equation}        
None of these have integer slopes.

We can now describe the possible test configurations. 
\begin{itemize}
	\item The possible candidates are associated to the subsets of polytopes obtained by forgetting one of the points $p_i$. In the discussion at the end of section \ref{ssub:pi}, this is the point placed at $z=\infty$, while all the other $p_i$ coincide in the $\lambda\to 0$ limit. 
  
		In our example, the candidate subsets would be three: one associated to the set of polytopes $\{\Delta_2, \Delta_3\}$ (obtained by forgetting $\Delta_1$, which is associated to $p_1$), one associated to the set of polytopes $\{\Delta_1, \Delta_3\}$, and finally one associated to $\{\Delta_1,\Delta_2\}$. 
	
	\item There is now a procedure to read off the toric diagram of the degeneration from each subset of $\Delta_i$. However, not all these candidates will produce valid test configurations: some will not be normal. For us it is enough to know when this happens. The criterion is as follows: a set of polytopes is called admissible only if \textit{at most one} of the $\Psi_i$ does not have integer slopes (recalling (\ref{eq:int-slope})). There is then a test configuration for each admissible candidate subset of $\Delta_i$.
	
In our example, none of the $\Psi_i$ has integer slopes. So none of our three candidates $\{\Delta_2, \Delta_3\}$, $\{\Delta_1, \Delta_3\}$ and $\{\Delta_1, \Delta_2\}$ is admissible, and we have no test configurations. This is in agreement with our naive analysis below (\ref{eq:p-uv-laufer}).
\end{itemize}

The test configurations obtained this way are the only ones that need to be checked for K-stability \cite{ilten-suss}. So in our example in this section there can be no test configurations, and we know already that the threefold is K-stable, and hence is a Calabi--Yau. 

In less lucky cases, there can be several test configurations. In principle \cite{ilten-suss} gives a way to produce explicitly the test configurations associated to an admissible collection of polytopes. However, as we mentioned, once one knows the number of test configurations, it is usually also easy to produce them explicitly by trial and error. 
 


\subsection{Examples: Brieskorn--Pham singularities} 
\label{sub:k-ex}

In \cite{collins-szekelyhidi}, the K-stability criterion was applied to three classes of singularity: in the Yau--Yu classification of all (hypersurface) singularities with at least one $\mathbb{C}^*$ action \cite{yau-yu}, these are the first three of nineteen. Here we briefly quote those results for the first class, the so-called Brieskorn--Pham $\mathrm{BP}(p,q)$ singularities:\footnote{As a curiosity, we note that quite a bit is known about homological mirror symmetry for these threefolds. For instance, the Fukaya category has been calculated in \cite{futaki-ueda}.}
\begin{equation}\label{eq:BP}
	uv + z^p + t^q=0\,.
\end{equation}
In section \ref{sub:yy123} we will look for quivers in this class, and also in the other two analyzed by \cite{collins-szekelyhidi}, YY-II and YY-III. We will find examples that are basically the ``generalized conifolds'' of $A_m$ type considered in \cite{gubser-nekrasov-shatashvili}.

For (\ref{eq:BP}), the isometry torus is $T=\mathrm{U}(1)^2$. The Reeb vector field $\xi$ can be found by volume minimization. We can then perturb by two test configurations and compute the Futaki invariant with respect to both, as explained in section \ref{sub:fut}:
\begin{equation}\label{eq:a0-BP}
	\xi = \frac{3}{2(p+q)} (pq,pq,2q,2p) \, ,\qquad a_0(\xi)= \frac{2}{27}\frac{(p+q)^3}{(pq)^2}\,.
\end{equation}
To find the number of test configurations, one can use the techniques of \cite{ilten-suss} explained in section \ref{sub:torus}. This time the base $B$ is described by an equation $w_0^m+ w_1^m+w_2=0$ in the weighted projective space $\text{W}\mathbb{CP}^{1,1,m}$, where $m=\mathrm{gcd}(p,q)$. This is still genus 0, but it intersects one of the toric divisors $m$ times rather than just one. Thus one of the three polytopes in the analogue of figure \ref{fig:pp} is now repeated $m$ times. Another difference with our example in that figure is that one of the polytopes $\Psi_1$ has integer slope. Among the candidate test configurations, $\{\Delta_1, \Delta_3\}$ and $\{\Delta_1, \Delta_2\}$ are then admissible, because in both cases only one of the $\Psi_i$ have non-integer slopes. This means that there are two test configurations in this case. Once we know that there are two of them, it is easy to find them more directly by hand:
\begin{align}
&\quad\lambda_1 = (0,0,1,0)\ , \quad \lambda_2 = (0,0,0,1)\ , \label{eq:fut-BP}\\  
&\quad \text{Fut}(\xi,\lambda_1) = \frac{1}{2}\left( \frac{2q-p}{3q}\right)a_0(\xi,0)\ , \quad \text{Fut}(\xi,\lambda_2) = \frac{1}{2}\left( \frac{2p-q}{3p}\right)a_0(\xi,0)\,. \nonumber
\end{align}
Imposing the positivity of the Futaki invariants, we see that $\mathrm{BP}(p,q)$ is K-stable if and only if
\begin{equation}\label{eq:stab-BP}
	1/2 < p/q < 2\,.
\end{equation}
For more details, see \cite[Sec.~8]{collins-szekelyhidi}.

In particular, $\mathrm{BP}(2,2)$ is nothing but the conifold (\ref{eq:con-eq}), which satisfies (\ref{eq:stab-BP}). On the other hand, $\mathrm{BP}(2,q)$ is stable for $q=3$ but for no other case. This is in agreement with the obstructions found in \cite{gauntlett-martelli-sparks-yau} for this class of examples.

\section{Quivers from matrix factorizations} 
\label{sec:nccr}

We will now explain how to use algebraic methods to extract quiver and superpotential of the gauge theory associated with the singularity. In section \ref{sub:nccr} we will introduce the notion of non-commutative crepant resolution (NCCR), which was already suggested in \cite{berenstein-leigh,aspinwall-morrison-quivers} as a physically relevant way to associate quivers to singularities. In \ref{sub:iw} we will describe an algorithm to find NCCRs for a certain class of singularities.

\subsection{Non-commutative crepant resolutions}
\label{sub:nccr}

Recall that in this paper we are restricting our attention to hypersurface singularities, i.e.~singularities defined by a single equation $p(u,v,z,t)=0$ in $\mathbb{C}^4$. We associate to it the ring
\begin{equation}
R \equiv \cc[u,v,z,t] / (p)\ ,
\end{equation}
namely, the ring of polynomials in $\mathbb{C}^4$, modulo an equivalence relation that sets $p$ to zero. 
We will require $R$ to be Gorenstein, namely that a holomorphic $(3,0)$-form exists (or in other words that the canonical bundle is trivial.) 

A resolution $\tilde Y$ is a non-singular space which is isomorphic to $Y$ almost everywhere; more precisely, there is a ``birational'' map $\tilde Y \dashrightarrow Y$, which induces an isomorphism from a nontrivial open set of $\tilde Y$ to one of $Y$. $\tilde Y \dashrightarrow Y$ is called \emph{crepant} if it does not change the canonical bundle; in our CY case, if $K_{\tilde Y}$ is trivial. A familiar example is where $Y$ is the conifold (\ref{eq:con-eq}), and $\tilde Y$ is the so-called resolved conifold, where the singularity is replaced by a $\mathbb{CP}^1$. 

Given the SCFT dual of a $Y=C(L_5)$, there is a branch of its moduli space corresponding to separating a single D3-brane from the others and moving it away from the singularity and along $Y$. This branch is obtained by taking certain small ranks 
\begin{equation}\label{eq:single-D3}
	N^{\mathrm{single\ D3}}_i
\end{equation}
in the quiver; often, but not always, one has $N^{\mathrm{single\ D3}}_i=1$. Since it corresponds to moving a D3-brane along $Y$, this branch should be isomorphic to $Y$ itself. If one introduces non-zero Fayet--Iliopoulos parameters in this small-rank quiver, one often obtains a crepant resolution $\tilde Y$.

Recall that a module $M$ over $R$ is an abelian group with an action $\cdot: R\times M \to M$ which is associative and distributive; one can think of it as a sort of ``representation'' of a ring (and indeed group representations are sometimes also called modules). If a ring is the algebraic manifestation of a manifold, a module is the algebraic representation of a bundle. 

The $R$-modules we will be interested in are called (maximal) Cohen--Macaulay modules (CM).\footnote{Much of the background material on CMs can be found in \cite{leuschke,wemyss-lec}.} Every module $M$ has a \emph{projective resolution}, namely it fits in a sequence 
\begin{equation}\label{eq:proj-res}
	\ldots \longrightarrow R^{\oplus n_2} \longrightarrow R^{\oplus n_1} \longrightarrow M \longrightarrow 0\,,
\end{equation}
which is exact, i.e.~the kernel of every map coincides with the image of the previous one. The maximal length of such a resolution is called the \emph{global dimension} of $R$, and gives an algebraic analog of the geometrical dimension. For a singular manifold, the global dimension is infinite: there are some $M$ whose projective resolution is infinitely long. But for such an $M$, there is a particular, eventually two-periodic, resolution:
\begin{equation}\label{eq:mfCM}
\begin{tikzcd}
\ldots \rar & R^{\oplus n} \rar{\Phi} & R^{\oplus n} \rar{\Psi}& R^{\oplus n} \rar{\Phi} & R^{\oplus n} \rar{\Psi} & R^{\oplus n} \rar{m} & M \rar &0\ ;
\end{tikzcd}
\end{equation}
exactness of the sequence implies 
\begin{equation}\label{eq:mf}
\Psi \cdot \Phi = \Phi \cdot \Psi = p\, 1_{n \times n}\ .
\end{equation}
Such a pair $(\Phi,\Psi)_n$ is called a \emph{matrix factorization} (MF) of the polynomial $p$. (The subscript denotes the matrix dimension.) In other words, $M$ is the cokernel of an $n\times n$ matrix $\Psi$ for which a $\Psi$ exists satisfying (\ref{eq:mf}).

We will most often present the relevant CM modules without writing down the explicit matrices $\Psi$ and $\Phi$. However, once one specifies $M$ and its generators, it is always possible to explicitly write down such matrices. The $n=1$ factorizations $(1,p)_1$ and $(p,1)_1$ always exist, and are referred to as trivial and non-reduced MF respectively. Any MF containing $(p,1)_1$ as a summand in a direct sum (i.e.~as a block) is said to be non-reduced. Only affine singular varieties (defined by $p=0$) admit a reduced, nontrivial MFs, which in favorable situations can often be classified. 

For example, for the conifold (\ref{eq:con-eq}), a non-trivial MF is given by
\begin{equation}\label{eq:mf-con}
	\Psi= \left( \begin{array}{cc}
		v & -f_1 \\
		f_2 & u
	\end{array}\right)
	\, ,\qquad
	\Phi= \left( \begin{array}{cc}
		u & f_1 \\
		-f_2 & v
	\end{array}\right)\,,
\end{equation}
where $u=x+iy$, $v=x-iy$, $f_1 \equiv z+ i t$, $f_2 \equiv z-it$, so that $p=uv + f_1 f_2$.
The cokernel of $\Psi$ is then a CM. Explicitly, this cokernel is generated by the vector $(u,f_1)$. In other words, the last map $R^{\oplus 2} \xrightarrow{m} M$ in (\ref{eq:mfCM}) is given by multiplication by $(v,- z)$; the composition $m \circ \Psi = (u, f_1)\left(\begin{smallmatrix}v & -f_1 \\ f_2 & u\end{smallmatrix}\right)= (p,0)=(0,0)$, as appropriate for an exact sequence. Still more explicitly, 
 $(\phi_1,\phi_2) \xrightarrow{m} u \phi_1 + f_1 \phi_2$; the module $M$ consists of all functions that can be written in this form (see footnote \ref{foot:ideal}). 

The general theory behind MFs has first been developed in \cite{eisenbud}, and they already made their appearance in physics in various contexts \cite{kapustin-li,baumgartl-bruner-gaberdiel,brunner-herbst-lerche-walcher,hori-walcher,gukov-walcher,brunner-herbst-lerche-scheuner,brunner-gaberdiel-keller,Caldararu:2007tc,herbst-hori-page,aspinwall-morrison-quivers,bcv-oldflux,Addington:2012zv,Sharpe:2012ji,collinucci-savelli-MF,collinucci-savelli-T,collinucci-fazzi-valandro,collinucci-fazzi-morrison-valandro}.

Once one has found a set of CM modules, one can define the non-commutative ring
\begin{equation}\label{eq:NCCR}
A \equiv  \End_R \left(R \oplus \bigoplus_i M_i \right)\ .
\end{equation}
In practice, $A$ can be presented as a quiver with relations, where each CM $M_i$ (we also put $R=M_0$ by convention) corresponds to a node, and maps between nodes are generated by arrows satisfying certain relations. We will see several examples below; some more can be found in e.g.~\cite{aspinwall-morrison-quivers} and \cite[Sec.~2]{collinucci-fazzi-valandro}.

(\ref{eq:NCCR}) is a generalization of $R$ itself, in the sense that for $n=0$ we have $ \End_R R \cong R$. If we include enough CMs in (\ref{eq:NCCR}), then it might happen that the global dimension of $A$ (the maximal length of projective resolutions (\ref{eq:proj-res}) over it) becomes finite, even though the one of $R$ was infinite. If this happens, and if moreover $A$ itself is a CM over $R$, $A$ is called a \emph{non-commutative crepant resolution} (NCCR) of $R$ \cite{vdb-nccr}.

For example, for the conifold the CM module $M_1$ defined by the MF in (\ref{eq:mf-con}) is already enough (together with $R=\mathbb{C}[x,y,z,t]/(x^2+y^2+z^2+t^2)$ itself) to produce an NCCR $\End_R ( R \oplus M_1)$. The two summands can be represented by two nodes. The endomorphisms are generated by two maps from $R$ to $M_1$ and two from $M_1$ to $R$; we will discuss these in detail in section \ref{subsub:KW}. The upshot is that one reproduces this way the familiar Klebanov--Witten quiver \cite{klebanov-witten}, as we will see in figure \ref{fig:quivm1}.

As the name implies, an NCCR is an algebraic analogue of a crepant resolution. Indeed for our case of dimension three, an NCCR guarantees the existence of a crepant resolution. (The assumptions that $A$ should be CM and have global dimension three are the non-commutative counterpart of $Y\dashrightarrow P$ being smooth and crepant.) Moreover, the NCCR and the crepant resolution have equivalent (bounded) derived categories, which have been suggested \cite{douglas,berenstein-leigh} to be the mathematical description of topological B-branes, the counterpart in the topological string of D-branes.

All this suggests that an NCCR gives a way to find the SCFT dual to a CY singularity, as suggested in \cite{berenstein-leigh,aspinwall-morrison-quivers}.  The ranks of the CM modules should correspond to the ranks $N^{\mathrm{single\ D3}}_i$ of the single D3-brane moduli space discussed around (\ref{eq:single-D3}). A large-$N$ generalization can then be obtained by taking ranks $N \times N^{\mathrm{single\ D3}}_i$. Several checks of this conjecture have already been carried out; we have mentioned that the conifold quiver is correctly reproduced, and so are for example the quivers for the (infinite class of) $Y^{p,q}$ metrics \cite{beil}. In this paper we will carry out several more such checks.

\subsection{\texorpdfstring{An algorithm for compound $A_m$ Du Val threefolds}{An algorithm for compound Am Du Val threefolds}}
\label{sub:iw}

A singularity of the type
\begin{equation}\label{eq:lift}
	0=p(x,y,z,t)= x^2 + y^2 + f(z,t) = uv + f(z,t)\,,
\end{equation}
where $u=x+iy$, $v=x-iy$, is called a \textit{lift} of the onefold singularity $f(z,t)=0$, since many of the properties of the threefold singularity are simply inherited from those of the onefolds.\footnote{For a list of rigorous results on lifts see \cite[Chap. 12]{yoshino}. (See also \cite[Sec. 2.3]{collinucci-savelli-MF} for a physics perspective in a different context.) E.g. one can prove that simple singularities (i.e. those without complex moduli) are of finite representation type in any dimension. For us, this means that the set of CMs is finite.}

In this section we will review an algorithm due to Iyama and Wemyss (IW) \cite[Sec.~5]{iyama-wemyss-cAn} that produces NCCRs $A$ for a certain class of $f$. The result will be similar to the ``generalized conifolds'' of $A_m$ type considered in \cite{gubser-nekrasov-shatashvili}; we will comment on the relation in section \ref{ssub:gen-con}.

\subsubsection{Algorithm for the quiver} 
\label{ssub:iw}

A hypersurface is a \emph{compound Du Val singularity} \cite{reid} of type $A_m$, or $cA_m$ for short, if its intersection with a generic hyperplane in $\cc^4$ is an $A_m$ surface (i.e. twofold) singularity. One can prove \cite[Prop. 6.1]{burban-iyama-keller-reiten} that any $cA_m$ threefold can be put into the form (\ref{eq:lift}), namely $uv+f(z,t)=0$, with $f(z,t)$ containing at most $m+1$ irreducible (prime) factors $f_i$ in a power series expansion (around the singular point), all of them vanishing at order $\text{ord} f_i=1$. 
Intuitively, this means $uv +f(z,t) \sim uv + z^{m+1} + \ldots$, which determines the integer $m$. Then, by \cite[Thm. 5.7]{burban-iyama-keller-reiten}, containing exactly $m+1$ factors is equivalent to the threefold admitting an NCCR.

Thus, for a $cA_m$ singularity, we simply need to check whether the polynomial $f(z,t)$ in $uv+f(z,t)=0$ can be factored into $n=m+1$ prime terms:
\begin{equation}\label{eq:nccr-crit}
\begin{split}
	Y:&\ p=uv + f = 0\quad \text{with} \quad f=f_1 \cdots f_{n=m+1} \, ,\\ 
	&f_i \in \mathfrak{m}\equiv (z,t)\ , \quad f_i \not\in \mathfrak{m}^2\ \forall i\ .    
\end{split}
\end{equation}
If this holds, then the singularity $R$ admits an NCCR. If it does not, there exists no NCCR. Here $\mathfrak{m}$ is the maximal ideal of the ring $S\equiv \cc[z,t]$, namely the ideal of linear functions, and $f_1 \cdots f_n$ is a factorization of $f$ (into prime elements of $S$). $\mathfrak{m}^2$ is then the ideal of quadratic functions; thus (\ref{eq:nccr-crit}) requires $f_i$ not to have a critical point at the origin.

\begin{figure}[!ht]
    \centering
    \begin{subfigure}[t]{.35\textwidth}
        \centering
       \includegraphics[scale=1.1]{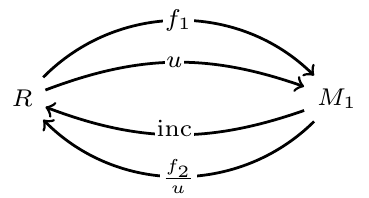}
\caption{Quiver for $m=1$, i.e. $n=2$.}
\label{fig:quivm1}
    \end{subfigure}%
~
    \begin{subfigure}[t]{.5\textwidth}
        \centering
       \includegraphics[scale=1.1]{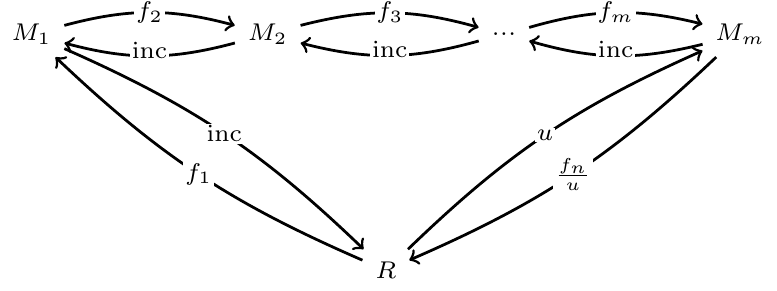}
\caption{Quiver for $m>1$, i.e. $n>2$.}
\label{fig:quivm2}
    \end{subfigure}
    \caption{The quivers presenting the NCCR $A=\End_R T$ \eqref{eq:NCCRmaxflag}. One may also have to add loops at vertices according to the rules in the main text. Since $\Hom_R(M_j,M_j)\cong R$ for $j=0,\ldots,m$ (with $M_0\equiv R$), we must see the generators of $R$ at every vertex. Indeed $v$ ($u$) is the complete (anti)clockwise loop based at $M_j$; $z,t$ can be seen by following the above rules.}
        \label{fig:quivnccr}
\end{figure}

If (\ref{eq:nccr-crit}) holds, the special set of CMs is constructed in terms of ideals of $R$ as follows:\footnote{\label{foot:ideal}The ideal $(g_1,\ldots,g_n)$ is defined as being the space of linear combinations $\sum a_i g_i $, with $a_i$ elements of the ring. So for example $(z,t)$ is the ideal of all functions vanishing at the origin, also called the ``maximal ideal'' of the origin. Notice that every ideal of $R$ is also a module of $R$: more precisely, ideals are the submodules of $R$, seen as a module over itself.}
\begin{equation}\label{eq:NCCRmaxflag}
T = \bigoplus_{j=0}^m M_j\ , \quad M_j \equiv \left(u, \prod_{i=1}^j f_i \right)\ .
\end{equation}
The quiver then is as in figure \ref{fig:quivnccr} \cite[Cor. 5.33]{iyama-wemyss-cAn}. One may have to add loops at each vertex according to the following rules:
\begin{itemize}
\item at vertex $R$, if $(f_1,f_n)=(z,t)$, add no loops. If $(f_1,f_n,e_0)=(z,t)$ for some element $e_0 \in \cc[z,t]$, add a loop at $R$ amounting to multiplication by $e_0$ in the ring. If such an element cannot be found, add two loops at $R$ amounting to multiplication by $z$ and $t$ respectively.
\item at each vertex $M_j$, if $(f_j, f_{j+1})=(z,t)$ add no loops. Conversely add the loop $e_j$ if $(f_j,f_{j+1},e_j)=(z,t)$ or add two loops $z$ and $t$ if no such $e_j$ can be found.%
\end{itemize}%
Now the relations among the arrows in the quiver. To find them, we have to keep in mind that the quiver furnishes a presentation of the endomorphism ring $A$ where each arrow (or path, that is a logical concatenation of arrows) amounts to multiplication by say $f_i$, or $u$, or a polynomial $f_i \cdots f_j$ (and ``inc'' simply means multiplication by $1$). Then $z$ and $t$ must commute as generators of the (commutative) polynomial ring $\cc[z,t]$, therefore they must also commute as paths in the quiver, if two paths $\alpha$ and $\beta$ amounting to multiplication by $z$ and $t$ respectively can be found. Therefore $zt=tz$ gives an abstract relation $\alpha \beta = \beta \alpha$ among arrows (which do not in general commute, being $A$ non-commutative), and e.g. $zt^2=t^2z$ would give another (again, if a path amounting to multiplication by $t^2$ can be found). Notice however that producing relations is far from being algorithmic, and often other relations can be obtained that greatly simplify those found by the above method. (This is akin to the problem of finding a superpotential given its F-terms.)

Notice that in this way one can in fact construct $n!$ quivers, depending on the ordering of the $f_i$.\footnote{There is a further generalization \cite[Sec.~5]{iyama-wemyss-cAn} where one considers a \textit{flag} $\mathcal{F}:\emptyset \equiv I_0 \subsetneq I_1 \subsetneq \ldots \subsetneq I_m \subsetneq I_{m+1}\equiv\{1,\ldots,n\}$, namely sets of elements in $\{1,\ldots,n\}$ of increasing sizes, defines functions $f_{I_j} \equiv \prod_{i\in I_j} f_i$, $g_j \equiv \frac{f_{I_j}}{f_{I_{j-1}}}$, and uses these $g_j$ to define a smaller set of CMs. These smaller quivers capture the geometry of partial resolutions (i.e. we only consider \emph{maximal flags}, for which $n=m+1$); they will not play a role for us.}

The fact that the quivers in figure \ref{fig:quivnccr} are of the affine $A$ type (apart from the possible presence of adjoints) should not surprise the reader, given our assumption on the $cA_m$ nature of the threefold. 

Finally, note that if (\ref{eq:nccr-crit}) does not hold, even if there is no NCCR, another notion has been introduced, that of maximal modification algebra (MMA). This can be thought of as the non-commutative counterpart of a resolution where the space has been resolved as much as possible: the remaining singularities are $\mathbb{Q}$-factorial terminal, which implies that the resolving them will change the canonical bundle. In appendix \ref{sec:mma} we consider this concept a bit further, and show in an example that it does not lead to SCFTs, as one might expect.


\subsubsection{Conifold}
\label{subsub:KW}

We will now illustrate the above IW algorithm with the conifold, whose dual SCFT is well-known \cite{klebanov-witten}. 
This theory was reproduced with matrix factorizations in \cite{aspinwall-morrison-quivers}; here we will use it to illustrate the more recent IW algorithm. The relevant quiver has indeed already appeared in figure \ref{fig:quivm1}. 

The polynomial $f$ factors into two prime factors, $f(z,t)=f_1 f_2 \equiv  (z+it)(z-it)$, so $n=2$ and $m=1$. Therefore the quiver has two nodes, $R$ and $M_1$, and maps:
\begin{equation}
\alpha_1\equiv  f_1\ ,\quad \alpha_2\equiv  u : R \to M_1\ ; \quad \beta_1\equiv  \text{inc}\ ,\quad \beta_2\equiv   f_2 / u: M_1 \to R\ .
\end{equation}
By ``inc'' we mean the inclusion of the ideal $(u,f_1)$ into the the ring $R$, i.e. the map $r u + s f_1 \mapsto r u + s f_1 \in R$. At the polynomial level, it is simply given by multiplication by one. There are no loops at the vertices, since the ideal $(f_1,f_2)=(z+i t, z-it)$ equals the maximal ideal $(z,t)$. As we anticipated, the quiver is the one in figure \ref{fig:quivm1}.

Now the relations. Given that the logical composition of paths gives an element of the commutative ring $R$, these paths must commute if they produce the same element. This gives a relation in the (abstract) non-commutative path algebra of the two-node quiver. For example, composing from left to right, or in other words simply multiplying the polynomials,
\begin{equation}
(z+it) \circ \text{inc} \circ u = u \circ \text{inc} \circ (z+it) \ \Leftrightarrow\ \alpha_1 \beta_1 \alpha_2 = \alpha_2 \beta_1 \alpha_1\ .
\end{equation}
In the same way we get $\beta_1 \alpha_i \beta_2 = \beta_2 \alpha_i \beta_1$ and $\alpha_1 \beta_i \alpha_2 = \alpha_2 \beta_i \alpha_1$ for $i=1,2$. 

The surmise of \cite{aspinwall-morrison-quivers} is that the quiver with relations obtained in this way is the physics quiver that one should associate to the singularity. The ranks of the CM modules are both equal to one; as we stated at the end of section \ref{sub:nccr}, this indicates that for the single D3-brane moduli space one has to take $N^{\mathrm{single\ D3}}_i=1$. For more general choices of ranks, the $\alpha_i$ and $\beta_i$ are interpreted as matrices, and the relations can be derived from the famous Klebanov--Witten superpotential \cite{klebanov-witten}
\begin{equation}\label{eq:Wconi}
W_{\text{BP}(2,2)}=\mathrm{Tr}\left(\epsilon^{ij}\epsilon^{kl} \alpha_i \beta_k \alpha_j \beta_l\right) = \mathrm{Tr}\left(\alpha_1\beta_1\alpha_2\beta_2 - \alpha_1\beta_2 \alpha_2\beta_1\right)\ .
\end{equation}
Thus the NCCR method indeed reproduces the correct quiver and superpotential in this case.

In the language of section \ref{sub:k-ex}, the conifold is $\mathrm{BP}(2,2)$. For illustration purposes, we will now show briefly what happens for the generalization $\mathrm{BP}(2,2k)$, namely $x^2 + y^2 + z^2 + t^{2k}=0$ (which is known as Reid's pagoda). 
Unfortunately, this does not give rise to a superconformal theory, since (\ref{eq:stab-BP}) is not satisfied (and as already noticed back in \cite{corrado-halmagyi,gauntlett-martelli-sparks-yau}). In section \ref{sub:yy123} we will deal with $\mathrm{BP}(p,p)$, where (\ref{eq:stab-BP}) is satisfied and the NCCR methods also apply. 

\begin{figure}[!ht]
\centering
\includegraphics[scale=1.25]{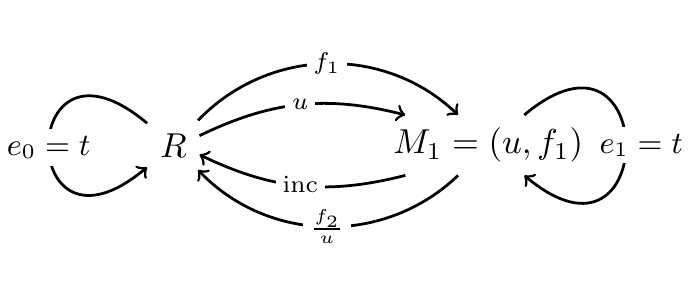}
\caption{Quiver with relations for BP$(p,q)=(2,2k)$, $k>1$ (i.e. Reid's pagoda \cite{reid}). $f_1=z+it^k$, $f_2=z-it^k$.}
\label{fig:reid}
\end{figure}

In the $\mathrm{BP}(2,2k)$ case, $f(z,t)=f_1 f_2 \equiv  (z+it^k)(z-it^k)$, $k>1$, so again $n=2$ and $m=1$. Now $f_1=z+it^k$ and $f_2=z-it^k$; the ideal $(f_1,f_2)$ is not equal to $(z,t)$, so we must add loops at both nodes. Adding the generator $t \in S$ does the trick, since clearly $(z+it^k,z-it^k,t)=(z,t)$; therefore on top of $\alpha_i , \beta_i$ we have a loop at $R$, call it $e_0$, and one at $M_1= (u, f_1)$, call it $e_1$, corresponding to multiplication by $t$ in the ring $S$. The quiver is depicted in figure \ref{fig:reid}.

We have two nontrivial relations
\begin{equation}
2it^k = (z+it^k)\circ  \text{inc} - u\frac{z-it^k}{u}\ \Leftrightarrow \ 2i e_0^k = \alpha_1 \beta_1 - \alpha_2 \beta_2\ ,
\end{equation}
and similarly $2i e_1^k = \beta_1 \alpha_1- \beta_2\alpha_2$ coming from paths $R\to M_1$ and $M_1\to R$ respectively, and four trivial ones such as $t(z+it^k)=(z+it^k)t$, implying $e_0 \alpha_i = \alpha_i e_1$ and the same with $\alpha_i \leftrightarrow \beta_i$. All of these can easily be integrated to a superpotential, namely \cite{aspinwall-katz,cachazo-fiol-intriligator-katz-vafa}: 
\begin{equation}\label{eq:Wreid}
W_{\text{BP}(2,2k)} = \mathrm{Tr}\left(\frac{2i}{k+1}e_0^{k+1}- \frac{2i}{k+1}e_1^{k+1} - e_0(\alpha_1\beta_1 - \alpha_2\beta_2) + e_1(\beta_1\alpha_1 - \beta_2\alpha_2)\right)\ .
\end{equation}
Notice that, for $k=1$, the superpotential terms $e_0^2$ and $e_1^2$ are masses for the adjoint fields; therefore we can integrate those out. Doing so lands us back on the conifold superpotential \eqref{eq:Wconi}. Once again we stress however that BP$(2,2k)$ is not superconformal.

\subsubsection{Relation to generalized conifolds} 
\label{ssub:gen-con}

The quivers in figure \ref{fig:quivm2} might remind the reader of the so-called ``generalized conifolds'' discussed in \cite{gubser-nekrasov-shatashvili}. We will comment here about the relation to that analysis. 

The generalized conifolds were obtained in \cite{gubser-nekrasov-shatashvili} by considering the quiver in figure \ref{fig:quivm2}, with adjoints $\Phi_i$ at every node. 
\begin{figure}[!ht]
\centering
\includegraphics[scale=1.25]{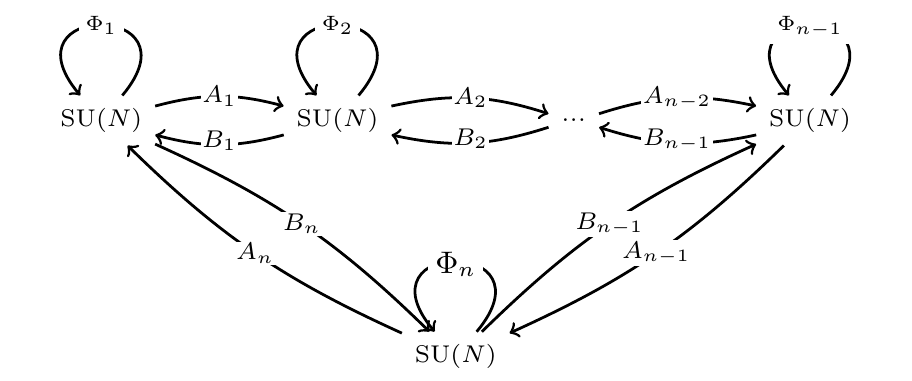}
\caption{The generalized conifold quiver in \cite{gubser-nekrasov-shatashvili}.}
\label{fig:gen-con}
\end{figure}
This quiver is originally obtained by the orbifold procedure \cite{douglas-moore} applied to $\mathbb{C}^2/\mathbb{Z}_n \times \mathbb{C}$, but in \cite{gubser-nekrasov-shatashvili} it is modified by adding a mass term $\Tr\Phi_i^2$ for each $i$ to the superpotential. Let us generalize their analysis slightly by turning these into more general functions $\Tr W_i(\Phi_i)$; we will call these ``higher-degree generalized conifolds''. In total the superpotential reads
\begin{equation}\label{eq:W-gen-con}
	W=\sum_{i=1}^n\mathrm{Tr}\left(\Phi_i (A_i B_i - B_{i-1} A_{i-1}) + W_i(\Phi_i)\right)\,.
\end{equation}
The ranks of the $\SU(N_i)$ gauge groups are all equal for conformality, while they can of course be kept different for more general fractional branes. 

One of the ways the dual geometry is identified in \cite{gubser-nekrasov-shatashvili} is by computing the abelian moduli space; the idea was described around (\ref{eq:single-D3}). This works as follows: we take all the ranks $N_i=1$, and the matrices $A_i$, $B_i$ and $\Phi_i$ all become complex numbers (which we denote by the corresponding lower-case letters $a_i$, $b_i$, $\phi_i$). Then the F-term equations obtained from (\ref{eq:W-gen-con}) read
\begin{equation}
	a_{i-1}b_{i-1}- a_i b_i = W_i'(\phi_i) \, ,\qquad b_i (\phi_i- \phi_{i+1})=0 = a_i (\phi_i - \phi_{i-1})\,,
\end{equation}
where a prime denotes differentiation w.r.t. $\phi_i$. The main branch of these equations is obtained by taking all $\phi_i=t$. Summing the first equation over $i$ then gives us the condition $\sum_i W_i'(\phi_i)=0$. Now if we define $u \equiv a_1 a_2 a_3$,  $y\equiv b_1 b_2 b_3$, $z\equiv a_1 b_1$, we get the equation $uv = \Pi_i (a_i b_i)$, or in other words 
\begin{equation}\label{eq:gen-con}
	uv = z \left(z+ W_2'(t)\right)\left(z+ W_2'(t)+ W_3'(t)\right)\cdots \left(z+\sum_{i=2}^n W_i'(t)\right)\,.
\end{equation}
This is the equation describing higher-degree generalized conifolds; if $W_i(\Phi_i)= \frac12 m_i \Phi_i^2$, then $W'_i(t)= m_i t$, which is the case originally considered in \cite{gubser-nekrasov-shatashvili}. 

While the abelian moduli space is not a particularly strong check of the proposed duality between (\ref{eq:W-gen-con}) and (\ref{eq:gen-con}), we will now see that the NCCR method confirms this proposal.

We can apply the IW algorithm of section \ref{ssub:iw} to (\ref{eq:gen-con}) simply by taking $f_1=z$, $f_2= z+ W_2'(t)$, \ldots, $f_n= z+ \sum_{i=2}^n W_i'(t)$. The equation is of compound $A_{n-1}$ type; so the IW algorithm tells us that indeed an NCCR exists, with the quiver in figure \ref{fig:quivm2}. As the caption there reminds us, we have to work out the possible existence of adjoints. Now we have two cases: 
\begin{itemize}
	\item If the polynomial $W_i$ is of degree 2 (the case in \cite{gubser-nekrasov-shatashvili}), the ideals $(W_i'(t),W_{i+1}'(t))$ are equivalent to the maximal ideal $(z,t)$, as one can see by linear combinations, so we need not add adjoints. The superpotential turns out to be the one we obtain from (\ref{eq:W-gen-con}) after integrating out the $\Phi_i$.
	\item If the $W_i$ have higher degree, the ideals $(W_i'(t),W_{i+1}'(t))$ do not include $t$ and thus are not equivalent to $(z,t)$. So we need to add adjoints at each node. The superpotential turns out to be (\ref{eq:W-gen-con}). 
\end{itemize}
To be sure, there are many cases which are covered by the IW algorithm but are \textit{not} of the higher-degree generalized conifold form (\ref{eq:gen-con}). The condition (\ref{eq:nccr-crit}) for an NCCR is met roughly speaking if the $f_i$ are linear in at least one variable, while (\ref{eq:gen-con}) requires all of them  to be linear in the same variable $z$.\footnote{In some cases, a linear change of variables might be needed to put a singularity in the form (\ref{eq:gen-con}); for example the suspended pinch point $uv=z^2t$ can be brought to the form (\ref{eq:gen-con}) by $(z,t)\mapsto (z,t+z)$. In this case actually a slightly different quiver can be used, which only has one adjoint \cite{butti-forcella-zaffaroni}; see also \cite{lopez}.} For example, for $p$ and $q$ even the equation (\ref{eq:Tpq}) below is a case where the IW algorithm gives an NCCR (with $f_1= z^{\frac{p-2}2}-t$, $f_2= z^{\frac{p-2}2}+t$, $f_3= z+ t^{\frac{q-2}2}$, $f_4= z- t^{\frac{q-2}2}$), but which cannot be written as (\ref{eq:gen-con}). 

To summarize: the original generalized conifolds of \cite{gubser-nekrasov-shatashvili} can be immediately generalized to the higher-degree form (\ref{eq:gen-con}). The IW algorithm covers cases which are still a bit more general, because the factors $f_i$ on the right-hand side of (\ref{eq:nccr-crit}) do not all have to be linear in the same variable; but it can still be seen as variations on the generalized conifold theme, so to speak.



\section{\texorpdfstring{K-stable $cA_m$ singularities}{K-stable cAm singularities}} 
\label{sec:cAm}

In this section, we will apply the IW algorithm of section \ref{sub:iw} to SE manifolds.  In section \ref{sub:yy123}, as a warm-up we look at the three classes of Sasaki--Einstein examples analyzed in \cite{collins-szekelyhidi}, one of which (the Brieskorn--Pham class) was reviewed in section \ref{sub:k-ex}. We will find that the existence of an NCCR on these SE manifolds puts stringent constraints, although it still leaves infinitely many cases. Given this, in section \ref{sub:gen} we follow a different approach and start directly from cases that have an NCCR, imposing K-stability later. This leaves us with a slightly larger class,
 
In section \ref{sub:minell} we change gears and look at ``minimally elliptic'' singularities, which are interesting as a generalization of the McKay correspondence.

\subsection{Yau--Yu classes I--III}
\label{sub:yy123}

\subsubsection*{Brieskorn--Pham (YY-I)} 

The first class in \cite{yau-yu} comprises the so-called Brieskorn--Pham manifolds $\mathrm{BP}(p,q)$; recall that the equation is $uv+ z^p+ t^q=0$.

According to the algorithm in section \ref{sub:iw}, we first have to ask whether the singularity is of compound type. It is easy to see that it is of $cA_{p-1}$ type (assuming $p \le q$). We then have to ask if $f(z,t)=z^p+t^q$ factorizes in $p$ factors. This only happens if $q/p \in \mathbb{N}$. However, (\ref{eq:stab-BP}) tells us that $q/p$ should be in the interval $(1/2,2)$. That leaves us with $q/p=1$ as the only choice. Thus in the following we will consider 
\begin{equation}
	\mathrm{BP}(p,p)\,.
\end{equation}

Assume $q=p\geq 2$ (for otherwise there is no singularity): we have to distinguish two cases according to the parity of $p$. Call $\omega \equiv e^{i\pi\, 2/p}$ a $p$-th root of unity. Then we have the following factorization into primes $f_i \in \mathfrak{m}$ (with $f_i \not\in \mathfrak{m}^2$):
\begin{equation}\label{eq:BPpp}
f=z^p+t^p=
\begin{cases}
\displaystyle \prod_{i=0}^{p-1} (z+\omega^i \,t)\ , & \text{$p$ odd;} \\
\displaystyle  \prod_{i=0}^{p-1} (z+\omega^{i+1/2}\, t)= \prod_{i=0}^{p-1} (z-e^{i\pi/p} \omega^i t) \ , & \text{$p$ even.}
\end{cases}
\end{equation}
Clearly $f_i=z+\omega^i t$ and $f_i=z+\omega^{i+1/2} t$ respectively. Observe that $(f_0,f_{p-1})=(f_i,f_{i+1})=(z,t)$ for $i=1,\ldots,p-2$. The CM modules are of the form $M_{i+1}=(u,\prod_{j=0}^{i} f_j)$ for $i=0,\ldots,p-2$. (By convention $M_0\equiv R$, whereas here $f_0\equiv z+t$ is nontrivial.) The arrows are $\alpha_i\equiv  f_i : M_{i} \to M_{i+1}$ and $\beta_i\equiv  \text{inc} : M_{i+1} \to M_i$ for $i=0,\ldots,p-2$. Finally, $\alpha_{p-1}\equiv  \tfrac{f_{p-1}}{u}$ and $\beta_{p-1} = u$, as usual. There are no loops at any node.
\begin{figure}[ht!]
        \centering
       \includegraphics[scale=1.1]{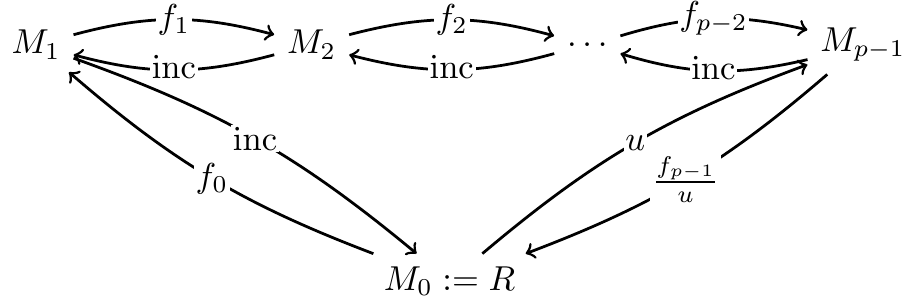}
    \caption{The NCCR of BP$(p,p)$.}
        \label{fig:quivBPpp}
\end{figure}
Given the cyclic structure of the quiver, the relations satisfied by the arrows can be assumed to come from a quartic superpotential of the form
\begin{equation}\label{eq:W-BPpp}
W_\text{BP$(p,p)$} = \sum_{i=0}^{p-1}\frac{1}{2} \mathrm{Tr}\left( A_i (\alpha_i\beta_i)^2 + s B_i \beta_i\beta_{i-1} \alpha_{i-1}\alpha_i\right)\ ,
\end{equation}
where the indices are $\mod p$. The two constants $A_i,B_i$ depend on $i$, whereas $s=\pm$ is a sign. The F-terms
\begin{align}
\partial_{\alpha_j} W &= A_j \beta_j \alpha_j \beta_j +s(B_j \beta_j \beta_{j-1}\alpha_{j-1} +B_{j+1}\alpha_{j+1}\beta_{j+1}\beta_j) \ , \\
\partial_{\beta_j} W &= A_j \alpha_j \beta_j\alpha_j + s(B_j \beta_{j-1}\alpha_{j-1}\alpha_j + B_{j+1}\alpha_j\alpha_{j+1}\beta_{j+1})\ ,
\end{align}
are then satisfied e.g.~by $A_j =(1+\omega)  \omega^{-j}$, $B_j= \omega^{-j}$, and $s=+$.

Given the quartic interactions, marginality of the superpotential imposes
\begin{equation}
R_{\alpha_i} +R_{\beta_i} =1 \ , \quad R_{\alpha_i} + R_{\alpha_{i-1}} +R_{\beta_i} +R_{\beta_{i-1}} =2\ .
\end{equation}
On the other hand, the beta function of each node automatically vanishes
\begin{equation}
\beta_i \equiv  N + \frac{N}{2}(R_{\alpha_i} + R_{\alpha_{i-1}} +R_{\beta_i} +R_{\beta_{i-1}} -1-1-1-1) = 0\ .
\end{equation}
Therefore the model (in the UV) is expected to flow to a fixed point, with the superconformal R-symmetry determined through $a$-maximization, which yields $R_{\alpha_i}=R_{\beta_i}=\tfrac{1}{2}$. The central charges then read
\begin{equation}
a=\frac{27p}{128} N^2-\frac{3p}{16}\ , \quad c=\frac{27 p}{128} N^2-\frac{p}{8}\ .
\end{equation}
As expected, $c-a=0$ at large $N$, and we can extract the following volume from $a$:
\begin{equation}
V(\xi) = \frac{N^2}{4a} = \frac{32}{27}\frac{1}{p}\ .
\end{equation}
This of course matches $2a_0$ in \eqref{eq:a0-BP} with $q=p$.

These theories are obtained from the generalized conifold of type $A_{p-1}$ \cite{gubser-nekrasov-shatashvili} by integrating out their massive adjoints \cite{butti-forcella-zaffaroni}.  Indeed notice that the defining equation of BP $(p,p)$ can be mapped to \cite[Eq. (6.10)]{butti-forcella-zaffaroni} (or \eqref{eq:gen-con}) via linear coordinate redefinitions. For example when $p$ is odd
\begin{equation}
z^p+t^p = \prod_{i=0}^{p-1} (z+\omega^i t)\ \longrightarrow\ \tilde{z}\tilde{t}(\tilde{z}+\tilde{t})\prod_{i=1}^{p-3}(\tilde{z}+\lambda_i\, \tilde{t})\ ,
\end{equation}
for appropriate $\lambda_i$.

\subsubsection*{YY-II} 
The second class in \cite{yau-yu}, as reviewed in section \ref{sub:k-ex}, consists of singularities with equation $uv+f(z,t)=0$, and $f(z,t)=z^p+zt^q$.

This is a compound $A_m$ with $m=\mathrm{min}(p-1,q)$. According to the criterion (\ref{eq:nccr-crit}) \cite[Sec.~5]{iyama-wemyss-cAn}, an NCCR exists if $f$ factorizes in $m-1$ factors. This is only the case if either $\frac{q}{p-1}\equiv r$ is an integer, or if its inverse $\frac{p-1}q \equiv s$ is an integer.

On the other hand their links will admit SE metric provided \cite[Sec.~8]{collins-szekelyhidi}
\begin{equation}\label{eq:stab-YYII}
	\frac{p^2-1}{2 p-1} < q<2 (p-1)\,
\end{equation}
or in other words $\frac{p+1}{2p-1}< \frac{q}{p-1}<2$. Since $\frac{p+1}{2p-1}>\frac12$ for any $p$, neither integer $r$ or $s$ defined above can be larger than one. So putting together the requirements for an NCCR and for an SE metric on the link we end up requiring 
\begin{equation}
	q= p-1\,,
\end{equation}
corresponding to the singularity
\begin{equation}
	uv + z^p + z t^{p-1}=0\,.
\end{equation}
Working out the algorithm of section \ref{sub:iw}, we get again the same quiver as in figure~\ref{fig:quivBPpp}. The superpotential is also the same as in (\ref{eq:W-BPpp}), but with different coefficients. 

A notable particular case is 
\begin{equation}
		\label{eq:D4} uv+z^3+ z t^2=0 \ ,\\
\end{equation}
which is an ``ADE threefold''. Such threefolds are close relatives of the perhaps more familiar ADE Du Val twofolds, surface singularities which can also be obtained as orbifolds $\mathbb{C}^2/\Gamma$ for $\Gamma$ a subgroup of $\mathrm{SU}(2)$; the well-known McKay correspondence states that resolving those singularities results in a set of $\cc\pp^1$'s intersecting according to the extended Dynkin diagram of an ADE group. An ADE threefold is obtained from an ADE twofold by adding a single square, similar to the procedure (\ref{eq:lift}) of adding two squares to lift a onefold. (\ref{eq:D4}) can be obtained in this way, and is a ``$D_4$ threefold'' (of type $cA_2$).

If we modify (\ref{eq:D4}) to the very similar-looking $uv+z^3+ z t^3=0$, we still have an ADE threefold, this time for $E_7$. This is still in the YY-II class, and it satisfies (\ref{eq:stab-YYII}) (so it has an SE metric), but does not admit an NCCR, since it is a $cA_2$ singularity but $z^3+z t^3=z(z^2+t^3)$ has two factors, not three as the criterion (\ref{eq:nccr-crit}) would require. As we discuss in appendix \ref{sec:mma}, in this case one can define a generalization of an NCCR called \emph{maximal modification algebra} (MMA); however, the $E_7$ example does not lead to a superconformal theory, and thus demonstrates that the concept of MMA does not seem to be physically relevant.\footnote{\label{foot:mma}The $A_k$ threefolds are BP$(2,k+1)$; they admit NCCR (trivial MMA) for odd (even) $k$, and have an SE metric only for $k=2,3$. The $D_k$ threefolds are YY-II$(k-1,2)$; they admit NCCR (MMA) for even (odd) $k$, but only $k=4$ admits an SE metric. $E_6,E_8$ are respectively BP$(3,4)$, BP$(3,5)$, they admit trivial MMA and an SE metric. By trivial MMA we mean that $f(z,t)$ does not factor at all, and the quiver is of the form presented on the bottom of figure \ref{fig:workflow}.}

\subsubsection*{YY-III} 
Finally we look at YY-III singularities; we recall from section \ref{sub:k-ex} that the singularities are defined by $uv+f(z,t)=0$, with $z^p t+zt^q$ and $p>1$, $q>1$.

The singularity is a compound $A_m$ with $m=\mathrm{min}(p,q)$; again an NCCR exists if $f$ factorizes in $m-1$ factors. This is only the case if either $\frac{q-1}{p-1}\equiv r$ is an integer, or if its inverse $\frac{p-1}{q-1} \equiv s$ is an integer.

The link admits an SE metric if and only if
\begin{equation}\label{eq:stab-YYIII}
3 (p-1)^2 (q-1) > (p+q-2)(pq-2p+1) \, ,\quad 3(q-1)^2(p-1) > (p + q - 2)(pq - 2q + 1)\,.
\end{equation}
Let us first analyze the case where $\frac{q-1}{p-1}= r$ is an integer. The second condition in (\ref{eq:stab-YYII}) reads then $\frac{r+1}3<\frac{r(q-1)}{rq-1}$. The latter is always $\le 1$ (with equality only if $r=1$). This immediately implies $r\le 2$; in fact for $r=2$ the condition becomes $\frac{2(q-1)}{2q-1}>1$, which is impossible. So the only possibility is $r=1$. 

We then look at the case where $\frac{p-1}{q-1} = s$ is an integer. The first in (\ref{eq:stab-YYII}) then reads $s+1 < 3 \frac{q-1}{q-2+s}$. The latter is always $\le 3$ (with equality only if $s=1$). So we have $s\le 2$; but for $s=2$ the inequality becomes $3<3\frac{q-1}q$, which is impossible. So the only possibility is  in fact $s=1$.

Thus the NCCR and SE requirements together give
\begin{equation}
	q= p\,,
\end{equation}
corresponding to the singularity
\begin{equation}
	uv + z^p t + z t^p=0\,.
\end{equation}

Again the algorithm of section \ref{sub:iw} gives the quiver in figure~\ref{fig:quivBPpp}, with $W$ of the form (\ref{eq:W-BPpp}) with some coefficients. 

\subsection{A simple generalization} 
\label{sub:gen}

In the previous subsection we saw that the existence of an NCCR puts severe constraints on the parameter space allowed by K-stability. Thus in this subsection we change our approach and try the opposite. We start from the class of singularities
\begin{equation}\label{eq:gen}
	uv + z^a t^b \prod_{i=1}^{k-a-b} (z- \lambda_i t^p)=0\,,
\end{equation}
where an NCCR is guaranteed to exist by (\ref{eq:nccr-crit}). The $f_i$ are given by $a$ copies of $z$, $b$ copies of $t$, and the factors $(z- \lambda_i t^p)$. As we remarked in section \ref{sub:iw}, there are many quivers that can be written for this case, depending on the ordering of the $f_i$; they all have $k$ nodes, but differ by the number and positions of the adjoints. With the ordering we have just given, the first $a-1$ nodes have an adjoint (corresponding to multiplication by $t$); the $a$-th node has no adjoint; the next $b-1$ nodes again have an adjoint (corresponding to multiplication by $z$); the remaining nodes have no adjoints.

To analyze K-stability, we first need to know the number of test configurations. If we want to apply the methods in \cite{ilten-suss}, the analysis differs from that for the example in section \ref{sub:torus} as follows. (Let us assume for simplicity generic $a$, $b$, $k$, $p$.) The base $B$ is now a submanifold
\begin{equation}\label{eq:w-sub}
	w_2=w_0^a w_1^b \prod_{i=1}^{k-a-b} (w_0- \lambda_i w_1)=0
\end{equation}
of weighted projective space $\text{W}\mathbb{CP}^{1,p,N}$, where $N\equiv p(k-b)+b$; the vectors $v_i$ are $v_1={-p \choose -N}$, $v_2 = {0\choose 1}$, $v_3={1\choose 0}$. 
The function $\Psi_1$ is zero, while $\Psi_2$ and $\Psi_3$ both have integer slope.

In fact in this case it is clearer to use the spirit of the analysis in \cite{ilten-suss} without using the combinatorial data of the polytopes. Recall from the end of section \ref{ssub:pi} that the possible degenerations are associated with $\mathbb{C}^*$ actions on $B$, that make the $p_i$ coincide in groups. In our case the $p_i$ are the zeros of the right-hand side of (\ref{eq:w-sub}). There are $a$ of them at $\{w_0=w_2=0\}$, $b$ of them at $\{w_1=w_2=0\}$, and others at $\{ w_2=0,\, w_0=\lambda_i w_1\}$, some of which may be possibly repeated; call $m_i$ the number of times a $\lambda_i$ appears in the product in (\ref{eq:gen}). Now for example the $\mathbb{C}^*$ action
\begin{subequations}\label{eq:gen-act}
\begin{equation}
	(u,v,z,t)\to (\lambda^a_1 u, v, \lambda_1 z, t)
\end{equation}
leads when $\lambda_1\to 0$ to a degeneration where (\ref{eq:gen}) becomes $uv= z^a t^{N-a}$. This corresponds to the $p_i$ all coinciding at $\{w_1=w_2=0\}$, except the $a$ that were located at $\{w_0=w_2=0\}$, which remain there. There is a similar action
\begin{equation}
	(u,v,z,t)\to (u,\lambda^b_2 v, z, \lambda_2 t)
\end{equation} 
\end{subequations}
fixing instead the $b$ $p_i$ located at $\{w_1=w_2=0\}$. More generally one can define an action where one rescales $(u,v,z',t)\to (\lambda u, v, \lambda z',t)$, where $z'\equiv z + \lambda_i t^p$; this fixes one of the points at  $\{ w_2=0,\, w_0=\lambda_i w_1\}$. 

The Futaki invariants of (\ref{eq:gen-act}) read respectively
\begin{subequations}\label{eq:gen-fut}
\begin{align}
\text{Fut}(\xi,\lambda_{1}) &= \frac{(p+1)^3 (-p (a p+a-2 k p+k)-b (p-1) (2 p-1))}{81 p^2 (-pb+b+k p)^2}\ , \\  
\text{Fut}(\xi,\lambda_{2}) &= \frac{(p+1)^3 (b (p-4) p+b-k (p-2) p)}{81 p (-pb+b+k p)^2}\ .
\end{align}
\end{subequations}
The parenthesis in $\mathrm{Fut}(\xi,\lambda_1)$ is smaller than $(2p-1)(b(1-p) +kp)$; imposing that this should be positive then gives $\frac kb>\frac{p-1}p$, and hence $\frac kp>1$. On the other hand, the parenthesis in $\mathrm{Fut}(\xi,\lambda_2)$ implies $\frac kb<\frac{p^2-4p+1}{p(p-2)}\equiv f(p)$. For $p>2$, $f(p)<1$ and we have a contradiction. For $p=2$, we see directly that $\mathrm{Fut}(\xi,\lambda_2)\propto -3b<0$. So only
\begin{equation}
	p=1
\end{equation}
remains.   (\ref{eq:gen-fut}) now imply $-2a+k>0$, $-2b+k>0$ respectively. We still have the other potential actions mentioned below (\ref{eq:gen-act}), but they are in fact similar to the ones we have already analyzed: with $z\to z+ \lambda_i t$, equation (\ref{eq:gen}) remains of the same form, but with $a$ replaced by the multiplicity $m_i$ of $\lambda_i$. So we conclude that 
\begin{equation}
	k< 2 \min (a,b,m_i)\,.
\end{equation}


\subsection{Minimally elliptic threefolds}
\label{sub:minell}

In this section we will comment on a class of singularities which are part of the discussion in the previous subsection, and have some interesting geometry: they are an elliptic generalization of the McKay singularities.

In dimension two, elliptic singularities $P$ are those for which the arithmetic genus $p_\text{a}(P) =1$, but there is no upper bound on the geometric genus $p_\text{g}(P)$ \cite{nemethi}.\footnote{They were introduced in \cite{wagreich}, and are a classic field of study in singularity theory since then. See e.g. \cite{yau-ell,yau-almost}, or \cite{laufer-ell} and references therein.} For comparison, rational singularities (the Du Val ADE twofolds) have $p_\text{a}(P)=p_\text{g}(P)=0$ \cite{artin}. (Both genera are topological invariants of the resolution, i.e.~they can be deduced purely from its resolution graph.) As is well known, the resolution graphs of rational double points (the tree of intersecting $\cc\pp^1$'s) are given by the Dynkin diagrams of type ADE. The intersection matrix of the exceptional curves coincides with the Cartan matrix, which we can think of as an effect of the McKay correspondence. The resolution graphs of elliptic singularities allow for many more possibilities, and were classified by Wagreich \cite{wagreich} and Laufer \cite{laufer-ell}. Laufer also introduced the notion of minimally elliptic singularities, i.e.~those for which $p_\text{g}(P)=1$ (which are Gorenstein \cite{yau-ell}).\footnote{In this case the fundamental cycle of the resolution (topologically and analytically) coincides with the anti-canonical divisor. The elliptic double and triple points, together with the rational double points, are the only singularities in dimension two which are Gorenstein isolated hypersurfaces.} 

A particularly interesting example of minimally elliptic singularities are the simply elliptic ones (El$(n)$ in the language of \cite{laufer-ell,kahn}): the exceptional locus $E$ is a single smooth elliptic curve (as opposed to a tree of $\cc\pp^1$'s for rational singularities) with self-intersection $-n$, and the resolution is the total space of the (complex) line bundle $\mathcal{O}_E(-n)$. Another such case is provided by the cusp (Cu$(n)$), the resolution graph being a cycle of rational curves ($\cc\pp^1$'s) which intersect according to a few possible patterns.\footnote{See e.g. \cite[Sec.~1 \& Prop. 5.3]{drozd-greuel-kashuba}. The analogs of minimally elliptic singularities for curves are well-known, and correspond to singularities of modality $m=1$ (i.e. those that depend on one modulus). They can be found in \cite[Sec.~15.1]{arnold}.}

In dimension two the links $L_3$ of these singularities were also studied, see e.g. \cite{kasuya}. For simply elliptic singularities, $L_3$ is an $S^1$ bundle over $T^2$ (hence a Seifert manifold), whereas for cusps it is a $T^2$ bundle over $S^1$. Also, the CMs of minimally elliptic singularities were listed by \cite{kahn,kahn-phd}, and \cite{burban-iyama-keller-reiten} used this to produce an NCCR of the singularity.\footnote{It is actually known how to construct CMs for all minimally elliptic singularities \cite{drozd-greuel-kashuba}. However, besides the simply elliptic case and the cusp, all other singularities are of so-called wild type. In our language, we would need to add an infinite set of CMs to produce an NCCR. This allows us to restrict our attention to the former two cases only.}

We will construct threefolds from these twofolds by lifting them, namely by adding a single square \cite{burban-iyama-keller-reiten}:
\begin{equation}\label{eq:minell}
P_{T_{p,q,2,2}}:\ uv+\lambda z^2 t^2 + z^p + t^q = 0 \ \subset\ \cc^4\ ;\quad \lambda \in \cc\, \backslash\, \{0,1\}\ , \quad \frac{1}{p} +\frac{1}{q} \leq \frac{1}{2}\ .
\end{equation}
We have to treat separately the case where this inequality is saturated and the case where it is not. 

\begin{figure}[!ht]
    \centering
       \includegraphics[scale=1]{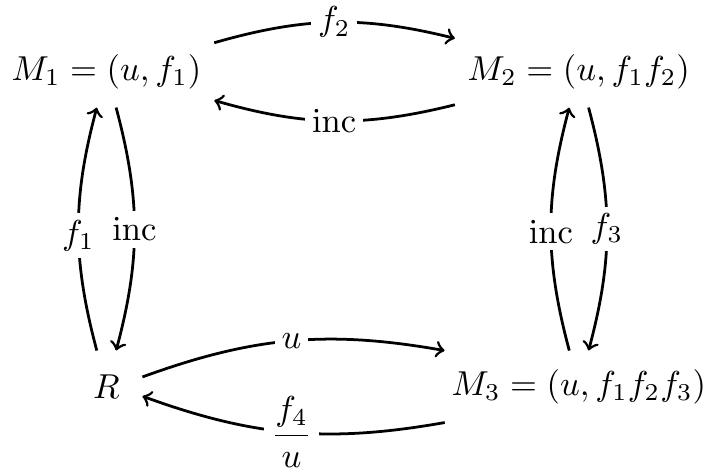}
\caption{NCCR for $P_{T_{4,4,2,2}(\lambda)}$. Here $f_1=t, f_2=z,f_3=z-t,f_4=z-\lambda t$.}
\label{fig:T4422}
\end{figure}

If the inequality in (\ref{eq:minell}) is saturated, it must be that $(p,q)=(3,6)$ or $(4,4)$: the corresponding twofolds are simply elliptic, while the threefolds admit the following equivalent presentations:
\begin{subequations}
\begin{align}
&P_{T_{3,6,2,2}(\lambda)}:\ uv+t(t-z^2)(t-\lambda z^2) = 0 \ \subset\ \cc^4\ ,\quad \lambda \in \cc\ , \label{eq:T36} \\
&P_{T_{4,4,2,2}(\lambda)}:\ uv+tz(z-t)(z-\lambda t) = 0 \ \subset\ \cc^4\ ,\quad \lambda \in \cc\  \label{eq:T44}
\end{align}
which can be obtained by factorizing $p(z,t)$ in (\ref{eq:minell}) and redefining $z,t$ and $\lambda$ appropriately.
Notice that the hypersurface equations depend on a complex modulus $\lambda$, hence the superpotential of the gauge theory obtained by having D3-branes probe $P_T$ will also depend on it. (Superpotentials with complex moduli have appeared previously, see e.g.~ \cite{wijnholt}.) 

If the inequality in (\ref{eq:minell}) is not saturated, $\lambda$ is unimportant (i.e.~one of the two coordinates can be shifted to reabsorb it), $(p,q)>(2,2)$, and the corresponding twofold is a cusp. The threefold hypersurface is
\begin{equation}
P_{T_{p,q,2,2}}:\ uv+(z^{p-2}-t^2)(z^2-t^{q-2})= 0 \ \subset\ \cc^4\ ; \quad \frac{1}{p} +\frac{1}{q} <\frac{1}{2}\ .\label{eq:Tpq}
\end{equation}
\end{subequations}
All these threefolds are $cA_m$ \cite{burban-iyama-keller-reiten}:
\begin{itemize}
\item $P_{T_{3,6,2,2}(\lambda)}$ is $cA_2$. We have $m+1=n=3$ prime factors, so an NCCR. As we can see by specializing the analysis in section \ref{sub:gen}, this singularity is however not K-stable.
\item $P_{T_{4,4,2,2}(\lambda)}$ is $cA_3$. We have $m+1=n=4$ prime factors, so an NCCR. Specializing section \ref{sub:gen} we see that this singularity is K-stable, with volume $\text{Vol}(L_5)=\frac{32}{27}\frac{1}{4}\pi^3$. We show the quiver for this singularity in figure \ref{fig:T4422}. The superpotential reads
\begin{multline}\label{eq:WT4422}
W_{T_{4,4,2,2}} = \frac{\lambda}{2}(\alpha_1\beta_1)^2 + \frac{1}{2}(\alpha_2\beta_2)^2 - \frac{\lambda}{2(1-\lambda)}(\alpha_3\beta_3)^2 - \frac{1}{1-\lambda}(\alpha_4\beta_4)^2  \ + \\ -\alpha_1\alpha_2\beta_2\beta_1 -\alpha_2\alpha_3\beta_3\beta_2+\alpha_4\alpha_1\beta_1\beta_4\ + \frac{1}{1-\lambda}\alpha_3\alpha_4\beta_4\beta_3\ .
\end{multline}
\item $P_{T_{p,q,2,2}}$ for $2(p+q)<pq$ is $cA_2$ for $p=3$ and $cA_3$ for $p>3$. We have $m+1=n=3,4$ prime factors respectively, hence an NCCR, if and only if $p=3$ and $q>6$ is even, or both $p,q>4$ are even. This case is of complexity two; so the techniques described in section \ref{sub:torus} do not apply, and we do not know how many test configurations we should expect. Even more worryingly, the Reeb vector would seem to be forced to be along the only $\mathrm{U}(1)$ action, which gives charges $(1,-1,0,0)$ to $(u,v,x,y)$. In view of this, we consider it unlikely that it gives rise to a Calabi--Yau threefold.
\end{itemize} 
The quivers can again be constructed by using the algorithm reviewed in section \ref{sub:iw}. 

\section{\texorpdfstring{Additional examples: compound $D_4$ threefolds}{Additional examples: compound D4 threefolds}} 
\label{sec:cD4}
   
So far we have found NCCRs by applying the IW algorithm of section \ref{sub:iw}. While this made it fast to find them, it has limited us to finding quivers which are morally similar to the higher-degree generalized quivers, as we commented in section \ref{ssub:gen-con}. In this section we break free of this limitation and explore more general cases. These are again of compound type, but to our knowledge no algorithm of the type in \ref{sub:iw} is available. We will reproduce two examples that were recently identified in \cite{amariti-fazzi-mekareeya-nedelin} by looking at the single-D3 moduli space, strengthening those dualities.

\subsection{A linear three-node quiver} 
\label{sub:3-n}

Consider the threefold
\begin{equation}\label{eq:cD4}
p=x^2 + ty^2 +t^2 z=x^2 + t(y^2+tz)=0\,.
\end{equation}
This singularity is not isolated: the gradient $dp$ vanishes along the entire $z$ axis. It is a compound $D_4$ singularity: for example if we intersect it with the non-generic hyperplane $y-z=0$ we get the $D_4$ equation $x^2 + yt(y+t)=0$.\footnote{The check with more generic hyperplanes $t=f(x,y,z)$ is more complicated. One way to establish it is of $cD_4$ type is to compute the Jacobi ring $\mathbb{C}[x,y,z]/\langle\partial_x p, \partial_y p, \partial_z p \rangle$, find a minimal set of generators by Gr\"obner bases methods, and compare with the generators of the Jacobi ring of the $D_4$ singularity. We did this by computer algebra.}

It is possible to resolve the singularity crepantly: above the origin (where the singularity is $cD_4$) we get a curve $\cc\pp^1_1 \cup \cc\pp^1_2$. Here the label on the $\cc\pp^1$ indicates its \emph{length}. (Thus $\cc\pp^1_1$ is an ordinary rational curve of genus zero and self-intersection $-2$, whereas by $\cc\pp^1_2$ we mean a length-two $\cc\pp^1$, which is an instance of non-reduced scheme. For more details see \cite{collinucci-fazzi-valandro,collinucci-fazzi-morrison-valandro}.) Above all other points along the $z$ axis (but the origin) the threefold is $cA_3$; upon resolving, we have three curves $\cc\pp^1_1 \cup \cc\pp^1_1 \cup \cc\pp^1_1$.

There are two $\mathbb{C}^*$ actions, acting on $(x,y,z,t)$ with the charge matrix
\begin{equation}\label{eq:charge-3n}
	\left(\begin{array}{cccc}
		1 & 0 & -2 & 2 \\
		0 & 1 &  4 &-2
	\end{array}\right)\,.
\end{equation}
The Reeb vector is given by a linear combination (with positive coefficients) of these two actions, $\xi = \kappa_1 \xi_1 + \kappa_2 \xi_2$, with $\xi_i$ generating the rows of (\ref{eq:charge-3n}). Volume minimization gives the Reeb vector
\begin{equation}
\xi = \left(\frac{3}{2} (\sqrt{3}+1),\frac{1}{2} (\sqrt{3}+3),2 \sqrt{3},3-\sqrt{3}\right)\,
\end{equation}
and the volume 
\begin{equation}\label{eq:volcase2geo}
\text{Vol}(L_5)= 2 a_0(\xi)\pi^3 = \frac{\pi^3}{3\sqrt{3}}\,.
\end{equation}

We now look at K-stability. Already at an intuitive level, we see that it is not easy to find test configurations: we can make (\ref{eq:cD4}) degenerate in various ways by making one of its monomials disappear in the central fiber $Y_0$, but this way we either obtain $x^2+t y^2=0$ or $x^2+t^2z=0$, which are copies of Whitney's umbrella, which is not normal as discussed around (\ref{eq:whitney}), or $t(y^2 +t z)=0$ which is not even irreducible. To make sure there are really no test configurations, we cause the algorithm in section \ref{sub:torus}, which works similar as to the example given in that section, and confirm the absence of test configurations.\footnote{Here are some details: the kernel of (\ref{eq:charge-3n}) is $\left(\begin{smallmatrix}-2&2&0&1\\ 0&-2&1&1\end{smallmatrix}\right)$; its columns give the fan of a singular toric space with four toric divisors $D_i$, and $B$ is a genus-zero curve inside it whose equation reads $1+X+XY=0$ in local coordinates $X\equiv \frac {ty^2}{x^2}$, $Y\equiv \frac{zt}{y^2}$, intersecting the $D_i$ in four points. The $\Psi_i$ are $\Psi_1=s/2$, $\Psi_2=t/2$, $\Psi_3=-s-t/2$ and $\Psi_4=-(s+t)/2$, none of which have integer slope.}  
Therefore a SE metric exists on the base of the CY$_3$ given by \eqref{eq:cD4}.

\begin{figure}[th!]
    \centering
       \includegraphics[scale=1.25]{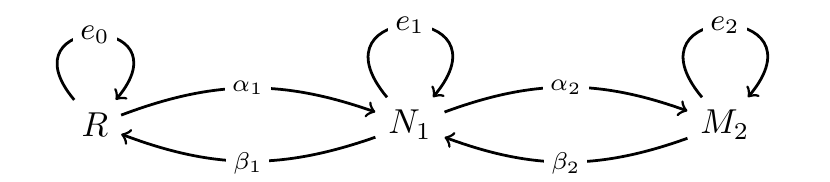}
\caption{The proposed NCCR for $R=\cc[x,y,z,t]/\text{\eqref{eq:cD4}}$. $N_1 \equiv \coker \Psi$ is a rank-two CM and corresponds to a physical $\SU(2N)$ group, whereas $M_2 \equiv \coker \psi$ is rank-one and corresponds to a physical $\SU(N)$ group.}
\label{fig:cD4nccr}
\end{figure}

Since the singularity \eqref{eq:cD4} is a $cD_4$ threefold, it is not of the form studied in section \ref{sub:iw}, which only applies to $cA_m$ threefolds, and we cannot use that algorithm. Thus in this case we simply look for the matrix factorizations (\ref{eq:mf}) by hand. We can take
\begin{subequations}
\begin{align}
	(\Phi,\Psi)_4 &= \left( \begin{bmatrix}  x & -y & -t & 0 \\ t y & x & 0 & -t \\  t z & 0 & x & y \\  0 & t z & -t y & x \end{bmatrix}, \begin{bmatrix}   x & y & t & 0 \\
	 -t y & x & 0 & t \\  -t z & 0 & x & -y \\  0 & -t z & t y & x \end{bmatrix} \right)\,; \\
(\phi,\psi)_2 &= \left( \begin{bmatrix}  x & -t \\  y^2+t z & x \end{bmatrix}, \begin{bmatrix}  x & t \\  -\left(y^2+t z\right) & x \end{bmatrix} \right) \,.
\end{align}
\end{subequations}
These two MFs define two CMs, respectively $N_1$, $M_2$ of rank two and one, via (\ref{eq:mfCM}); we then define an algebra $A$ via (\ref{eq:NCCR}). Recall that this $A$ is NCCR if it is Cohen--Macaulay and if its global dimension is finite. The check of the CM property can be done by computer \cite{singular}. Showing finite global dimension is in general difficult. However, following \cite{wemyss-hommmp}, one can argue  that there exists a unique rank-four CM generator $\Lambda$ such that $A=\End_R(\Lambda)$ is an NCCR. Since we have found one, namely $R\oplus N_1 \oplus M_2$, it must be that $\End_R(R \oplus N_1 \oplus M_2)$ is the NCCR we are after.\footnote{We would like to thank M.~Wemyss for discussions on this point.}  One can now compute the relations in the quiver and the superpotential using the prescription explained in \cite{aspinwall-morrison-quivers} (or via the path algebra procedure explained in \cite{karmazyn,collinucci-fazzi-morrison-valandro}). We get the quiver in figure \ref{fig:cD4nccr}, with superpotential 
\begin{equation}\label{eq:WcD4}
W = \Tr \left(e_0\, \alpha_1 \beta_1 + e_1^2 (\beta_1\alpha_1 +  \alpha_2\beta_2) + e_2\, \beta_2 \alpha_2 \right)\,.
\end{equation}
 
This quiver was already found to correspond to the singularity (\ref{eq:cD4}) by computing the single-D3 moduli space (\ref{eq:single-D3}). The ranks $N_i^{\text{single D3}}$ are the ranks of the CM modules in figure \ref{fig:cD4nccr}, namely $(1,2,1)$. The gauge invariants are given by \cite[Eq.~(D.53)]{amariti-fazzi-mekareeya-nedelin},
and satisfy the hypersurface equation \eqref{eq:cD4} upon imposing the F-terms coming from the superpotential \eqref{eq:WcD4}.

Marginality of the superpotential constrains the R-charges of the various fields, which we can parameterize via
\begin{equation}
R(e_1)=\Delta\ , \quad R(e_0)=R(e_2)=2\Delta\ , \quad R(\alpha_i)=R(\beta_i)=1-\Delta\,.
\end{equation}
In terms of $\Delta$ the central charges are given by
\begin{subequations}
\begin{align}
a(\Delta) &= \frac{27}{8} (\Delta -2) (\Delta -1) \Delta N^2 -\frac{3}{32} \Delta  \left(51 \Delta ^2-81 \Delta +40\right)\, , \\
c(\Delta) &= \frac{27}{8} (\Delta -2) (\Delta -1) \Delta N^2 + \frac{1}{32} \Delta  (9 (27-17 \Delta ) \Delta -110)\, .
\end{align}
\end{subequations}
As expected \cite{benvenuti-hanany}, they are equal at large $N$. Maximizing $a$ with respect to $\Delta$ we obtain the fixed-point value $\Delta_* = \tfrac{1}{3}(3-\sqrt{3})$, where $a$ attains the value $a=\tfrac{3}{4}\sqrt{3}N^2 +O(N^0)$. This means the dual $L_5$ has an SE metric with volume
\begin{equation}
\text{Vol}(L_5) = \frac{a_{{\mathcal N}=4\, \text{SYM}}}{a} \text{Vol}(S^5) = \frac{N^2}{4a(\Delta_*)} \pi^3= \frac{\pi^3}{3\sqrt{3}}\,,
\end{equation}
matching (\ref{eq:volcase2geo}).



\subsection{Laufer degeneration} 
\label{sub:uv-laufer}

We now consider the singularity 
\begin{equation}\label{eq:UVlaufer}
x^2+y^3+z^2 t = 0\,.
\end{equation}
It has featured recently in \cite{amariti-fazzi-mekareeya-nedelin}, and is a degeneration of the Laufer singularity (\ref{eq:laufer}) we will consider in the next section. 

It has two $\mathbb{C}^*$ actions, with a charge matrix we gave back in (\ref{eq:C*reeb-UV}),
on the coordinates $(x,y,z,t)$. The Reeb vector that minimizes the volume is given by
\begin{equation}\label{eq:Reeb-UVlaufer}
\xi_\text{UV-L} =\left( \frac{3}{10} (\sqrt{19}+7),\frac{1}{5} (\sqrt{19}+7),\frac{1}{2} (\sqrt{19}+1),\frac{2}{5} (8-\sqrt{19})\right)\,,
\end{equation}
leading to 
\begin{equation}\label{eq:futvolUV}
\text{Vol}(L_5) =\frac{1}{243} \left(19 \sqrt{19}-28\right)\pi^3\, .
\end{equation}

The counting of test configurations was performed already in section \ref{sub:torus} to illustrate the general procedure; it was concluded there that none are necessary. Thus (\ref{eq:UVlaufer}) gives rise to a Calabi--Yau threefold.

\begin{figure}[ht!]
\centering
\includegraphics[scale=1.5]{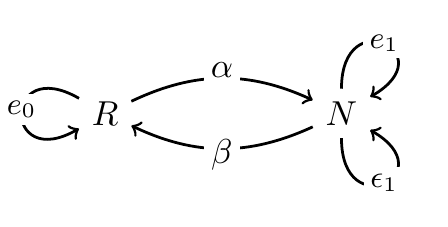}
\caption{The NCCR of $R=\cc[x,y,z,t]/\eqref{eq:UVlaufer}$. $N$ is a rank-two CM (corresponding to an $\SU(2N)$ gauge group), which can be obtained from the matrix factorization in (\ref{eq:MF-UVL}).}
\label{fig:UVlaufer}
\end{figure}

We now look for the quiver by using matrix factorizations. This can be done using techniques discussed in \cite{collinucci-fazzi-valandro,collinucci-fazzi-morrison-valandro},\footnote{\label{foot:UV}The hypersurface \eqref{eq:UVlaufer} can be obtained as a threefold slice of the so-called universal flop of length two \cite{curto-morrison}, i.e.~the sixfold $X^2 + U Y^2 +2 V Y Z + W Z^2+(UW-V^2) T^2 = 0 \subset \cc^7$, by taking e.g.~$X=x,\ Y=z,\ Z=y,\ U=t,\ V=0,\ W = y,\ T=0$. (See \cite{collinucci-fazzi-valandro,collinucci-fazzi-morrison-valandro} for more details.) Applying the cut to the MF of the universal flop we obtain (\ref{eq:MF-UVL})} and leads to 
\begin{equation}\label{eq:MF-UVL}
(\Phi,\Psi)_4 = \left( \begin{bmatrix}   x & -z & -y & 0 \\  t z & x & 0 & -y \\  y^2 & 0 & x & z \\  0 & y^2 & -t z & x \end{bmatrix}, \begin{bmatrix}    x & z & y & 0 \\  -t z & x & 0 & y \\ -y^2 & 0 & x & -z \\  0 & -y^2 & t z & x \end{bmatrix} \right)\,.
\end{equation}
As usual this defines a CM module $N$ via (\ref{eq:mfCM}), which has rank two. It turns out that $A=\mathrm{End}(R \oplus  N)$ is already an NCCR. It leads to the quiver in figure \ref{fig:UVlaufer}, with superpotential
\begin{equation}\label{eq:WUV}
W =\Tr \left(\beta e_0 \alpha +\alpha \epsilon_1^2\beta +\epsilon_1 e_1^2\right)\,.
\end{equation}
Again this was already obtained in \cite[Sec.~4.2]{amariti-fazzi-mekareeya-nedelin} by different methods.

As a cross-check we can again perform $a$-maximization. Doing so yields the IR R-charges \cite[Eq. (4.14)]{amariti-fazzi-mekareeya-nedelin}
\begin{equation}
	R(\alpha)=R(\beta)=1-\Delta_* \, ,\qquad R({e_0})=2\Delta_* \, ,\qquad R({\epsilon_1})=\Delta_* \, ,\qquad R({e_1})=1-\tfrac{\Delta_*}{2} \,
\end{equation}
with $\Delta_* =\tfrac{2}{15}(8-\sqrt{19})$. This agrees with the earlier result (\ref{eq:Reeb-UVlaufer}), once we take into account that the coordinates in (\ref{eq:UVlaufer}) are the gauge invariants
\begin{equation}\label{eq:UVL-gi}
	x = \alpha e_1 \epsilon_1 \beta  \, ,\qquad y = \alpha \epsilon_1 \beta \, ,\qquad
	z = \alpha e_1 \beta \, ,\qquad t = - \epsilon_1^2\,.
\end{equation}
The anomalies turn out to be
\begin{equation}\label{eq:a-UVL}
	\begin{split}
		a_\text{UV-L} &=\frac{1}{100} \left(19 \sqrt{19}+28\right) N^2 + \frac{1064 \sqrt{19}-5857}{3000}\,,\\
		c_\text{UV-L} &=\frac{1}{100} \left(19 \sqrt{19}+28\right) N^2 + \frac{2003 \sqrt{19}-10714}{6000}\,;
	\end{split}
\end{equation}
this $a$ is in agreement with \eqref{eq:futvolUV} via \eqref{eq:acharge-vol}.


\subsection{Laufer's theory} 
\label{sub:laufer}

We now turn to the Laufer singularity
\begin{equation}\label{eq:laufer}
	x^2+y^3+z^2 t+ y t^3=0\,.
\end{equation}               
The quiver was first constructed in \cite{aspinwall-morrison-quivers}, and has appeared recently in physics in \cite{collinucci-fazzi-valandro,amariti-fazzi-mekareeya-nedelin}. 

It has only one $\mathbb{C}^*$ action, which is given by the charge matrix 
\begin{equation}\label{eq:ai-laufer}
	(9,6,7,4)\,.
\end{equation}
Given that it is complexity two, we cannot apply the method reviewed in section \ref{sub:torus}. 
It is easy however to find some test configurations by hand. In the notation (\ref{eq:ti}):
\begin{equation}\label{eq:laufer-test}
	\lambda_{1}=(1,0,0,0) \, ,\qquad
	\lambda_{2}=(0,6,1,-2) \, ,\qquad
	\lambda_{3}=(0,0,1,0) \, ,\qquad
	\lambda_{4}=(0,0,-1,2) \, .
\end{equation}
These make (\ref{eq:laufer}) degenerate respectively to $y^3+z^2t+y t^3=0$, $x^2+z^2t + y t^3=0$, $x^2+y^3+yt^3=0$, and finally to $x^2+y^3+z^2t=0$, which is our old friend (\ref{eq:UVlaufer}). The Futaki is positive for all four: from (\ref{eq:exp-Fut}) we obtain that $\mathrm{Fut}/a_0$ is respectively $\frac{19}{27}$, $\frac{23}7$, $\frac13$, $\frac1{21}$. We do not have the general method of section \ref{sub:torus} to definitely make sure our test configurations are all that exist,\footnote{Other test configurations can be obtained by embedding (\ref{eq:laufer}) in $\mathbb{C}^d$ for $d>4$; we thank H.~S\"u\ss~for suggesting some examples.} but these preliminary checks  suggest the Laufer singularity (\ref{eq:laufer}) is a Calabi--Yau threefold.\footnote{A similar analysis can be performed for the generalization $x^2+y^3+z^2 t  + t^{2n+1}y=0$; however, already for $\lambda_{4}$ the Futaki is negative for $n>1$. It would be easy to repeat the calculation (also for the NCCR) for the model in \cite{pinkham,aspinwall-morrison-quivers}. The latter is a $cD_4$ threefold $p(x,y,z,t;\lambda)=0$ with an isolated singularity at the origin, and it depends on a complex modulus $\lambda$. When $\lambda=0$, it coincides with Laufer with $n=1$. However, for $\lambda\neq 0$ it is complexity-three, and we cannot use the method of section \ref{sub:torus} to check K-stability.}

The matrix factorization and quiver for Laufer, similar to (\ref{eq:MF-UVL}), is discussed at length in  \cite{aspinwall-morrison-quivers,collinucci-fazzi-valandro}. The quiver is the one in \cite[Fig.~6]{amariti-fazzi-mekareeya-nedelin}; it is similar to the one in figure \ref{fig:UVlaufer}, but without the adjoint $e_0$. The gauge invariants are still the ones in (\ref{eq:UVL-gi}). The superpotential can be obtained from (\ref{eq:WUV}) by adding a mass term $m e_0^2$ (and integrating out $e_0$), as well as a quartic deformation for $\epsilon_{1}$. This suggests the presence of an RG flow going from (\ref{eq:UVlaufer}) to (\ref{eq:laufer}) (similar to the one connecting $\mathbb{C}^2/\mathbb{Z}_2 \times \mathbb{C}$ to the conifold, which served as an illustration of test configurations back in section \ref{sub:fut}). 

Indeed in this case there is no $a$-maximization to perform, since there is only one $\mathbb{C}^*$ action; the $\alpha_i$ are given directly by (\ref{eq:ai-laufer}), and $a_0$ is given by (\ref{eq:a0-alpha}). From this (or \eqref{eq:a-4dquiv}) we read off the central charge:
\begin{equation}
a_\text{L} = \frac{567}{512} N^2 +O(N^0)\,,
\end{equation}
whose $N^2$ coefficient is smaller than the one in (\ref{eq:a-UVL}).

We remarked that $\lambda_{4}$ in (\ref{eq:laufer-test}) makes the Laufer singularity degenerate to (\ref{eq:C*reeb-UV}). Indeed $\lambda_{4}$ is one of the rows of (\ref{eq:C*reeb-UV}). One might then have the impression that the generalized $a$-maximization for (\ref{eq:laufer}) is in fact the same computation as ordinary $a$-maximization for (\ref{eq:UVlaufer}), since in both cases we vary with respect to the two $\mathrm{U}(1)$'s in (\ref{eq:C*reeb-UV}). This is however not the case, because in generalized $a$-maximization for the Laufer singularity (\ref{eq:laufer}) we are only allowed to add $\lambda_{4}$ with a positive coefficient $\epsilon$: since the Futaki invariant is positive, we do not lower $a_0$ by doing this. With ordinary $a$-maximization for the Laufer degeneration (\ref{eq:UVlaufer}), we do not have this restriction, and we can in fact find a lower value for $a_0$ by going in the negative $\epsilon$ direction; this is the minimum we gave earlier in (\ref{eq:Reeb-UVlaufer}).

It is perhaps clearer to rephrase this in terms of R-charges. For the Laufer model they read $R(\alpha)=R(\beta)=R({\epsilon_1})=\frac12$ and $R({e_1})=\frac 34$; the R-charges of the gauge invariants $(x,y,z,t)$ then become $\frac2{3b}(9,6,7,4)$, namely the appropriate normalization of the charge matrix (\ref{eq:ai-laufer}). Generalized $a$-maximization requires one to deform these: $R(\alpha)=R(\beta)\sim\frac12 - \delta$, $R({\epsilon_1})\sim\frac12 + \delta$, and $R({e_1})\sim\frac34-\frac \delta2$. This is in such a way that the gauge invariants $(x,y,z,t)$ get R-charges deformed by the test configuration $\lambda_{4}=(0,0,-1,2)$. This deformation of R-charges makes $a$ smaller for positive $\delta$. One would need to take negative $\delta$ to make $a$ smaller; but this is actually in contradiction with the hypothesis that the term $y t^3$ in the chiral ring equation should go to zero in the IR.



\section{Conclusions} 
\label{sec:conc}

In this paper, we have put together the techniques of non-commutative crepant resolutions (NCCR) and K-stability. The first deals more with the complex-geometry aspect of a singularity, while the second is a criterion for the existence of a Ricci-flat metric. 

While we have found several examples where the two can be put together and hence produce new holographic pairs, it is perhaps a little surprising that there are many more cases where only one of the two tests succeeds.

When an NCCR exists but K-stability fails, the canonical bundle is trivial and a quiver can be found, but there is no Ricci-flat metric. In fact in type IIB the general analysis of Minkowski flux vacua \cite{gmpt2} requires a complex structure (or more generally an odd generalized complex structure) with trivial canonical bundle, but not necessarily a compatible K\"ahler structure or a Ricci-flat metric. It might be that these singularities can then be used for holographic dualities involving fluxes; it would be rather interesting to explore this further. 

On the other hand, when K-stability succeeds but an NCCR does not exist, the situation is more puzzling. We have examined a more permissive version of NCCR which has been proposed in the mathematical literature, called maximal modification algebra (MMA), and unfortunately we have found that it does not produce SCFTs. So there appears to be no way to produce a physical quiver. One of the roles of a quiver in string theory is to describe fractional branes, but in cases without NCCRs there are also no crepant resolutions; perhaps fractional branes can only be defined when a crepant resolution exists. Another role of the quiver, however, is to produce SCFT duals to the Ricci-flat metric. Either there is a secret obstruction for some Calabi--Yau's to make sense in string theory, or for some singularities the SCFT is in fact non-Lagrangian. Clearly this is another point that requires more investigation in the future.

It would also be interesting to extend this paper to three-dimensional  ${\mathcal N}=2$ theories. K-stability techniques work pretty much in the same way. However, some of the results about NCCRs do change across dimensions; for example, an NCCR implies the existence of a crepant resolution only in dimension three. So the physical interpretation of an NCCR for M2-branes probing a fourfold might require further work before proceeding.

As we mentioned in section \ref{sub:phys}, the idea of K-stability seems to have a natural-enough field theory interpretation \cite{collins-xie-yau}, in terms of  degenerations of the chiral ring, which gives a concrete way of checking for emergent IR symmetries. A variant of this idea has already been considered beyond holography in \cite{benvenuti-giacomelli}, where terms in the superpotentials are dropped in the IR directly, without a direct reference to a $\mathbb{C}^*$ action. It would be interesting to compare the two procedures, and more broadly to see how well K-stability does in supersymmetric theories that do not have a string theory origin. Another way that the field theory interpretation might have an interesting interplay with geometry is in trying to restrict the number of test configurations that one has to check; \cite{xie-yau} recently tried to use the field theory interpretation to achieve this, and it might be interesting to see if there is any contact with the complexity-one procedure reviewed in section \ref{sub:torus}.

Finally there are a few obvious extensions of our methods to more general singularities. One direction is to consider complete intersection Calabi--Yau's (CICY), namely $n$ equations in $\mathbb{C}^{3+n}$, rather than the hypersurface ($n=1$) case we have considered here. The extension of the K-stability techniques is straightforward (indeed some cases already appeared in \cite{xie-yau}); moreover, the theory of matrix factorizations for CICYs exists already \cite{eisenbud-peeva,Eisenbud2016}.



\section*{Acknowledgments}
We would like to thank A.~Amariti, S.~Benvenuti, A.~Collinucci, N.~Ilten, N.~Mekareeya, S.~S.~Razamat, J.~Stoppa, H.~S\"u\ss, G.~Sz\'ekelyhidi and A.~Zaffaroni for interesting discussions. We are especially indebted to J.~Karmazyn and M.~Wemyss for numerous illuminating discussions. M.F.~would like to thank the 2014 University of Edinburgh ``Homological Interactions between Representation Theory and Singularity Theory'' workshop for a stimulating environment, and the  2019 Pollica summer workshop, which was supported in part by the Simons Foundation (Simons Collaboration on the Non-perturbative Bootstrap) and in part by the INFN, for hospitality during the final stages of this work. We would like to thank the Aspen Center for Physics, where this project was initially conceived, and each other's institution for hospitality at various stages. M.F.~acknowledges financial support from the Aspen Center for Physics through a Jacob Shaham Fellowship Fund gift.  The work of M.F.~is also partially supported by the Israel Science Foundation under Grant No.~504/13, 1696/15, 1390/17, and by the I-CORE Program of the Planning and Budgeting Committee. A.T.~is supported in part by INFN and by the ERC Starting Grant 637844-HBQFTNCER.


\appendix

\section{Maximal modification algebras} 
\label{sec:mma}

We have seen in section \ref{sub:iw} that a $cA_m$ singularity admits an NCCR if and only if (\ref{eq:nccr-crit}) applies with $n = m+1$. If $n \neq m+1$ the quiver in figure \ref{fig:quivnccr} does not provide an NCCR, but rather a so-called \emph{maximal modification algebra} (MMA) \cite{iyama-wemyss-AR,iyama-wemyss-Qf}. We will not need the precise definition of this object; suffice it to say that it is the non-commutative counterpart of a $\qq$-factorial terminalization, i.e.~a birational morphism $\tilde Y \dashrightarrow Y$ where $\tilde Y$ has at most $\qq$-factorial terminal singularities, which as we mentioned in the main text do not admit crepant resolutions.\footnote{In dimension three and over $\cc$ the existence of an NCCR $A = \End_R(R\oplus \bigoplus_i M_i)$ is equivalent to the existence of a crepant resolution $\tilde Y \dashrightarrow Y=\Spec R$ \cite[Thm. 6.6.3]{vdb-nccr}. The (singular) stable category $\underline{\text{CM}}(A)$ of CMs over $A$ \cite{yoshino} being zero means geometrical smoothness; the Cohen--Macaulay property, i.e.~$A \in  \underline{\text{CM}}(R)$, is instead the homological counterpart of crepancy. The existence of an MMA $A$ is equivalent to the existence of a $\qq$-factorial terminalization. This means that there can be points $y_i$ on $Y$ which are isolated hypersurface singularities, namely the localization $\mathcal{O}_{Y,y_i}$ of the structure sheaf of $Y$ at the (Zariski-closed point) $y_i$ is a hypersurface. ``$\qq$-factorial'' means that if $D$ is a Weil divisor, then $nD$ is Cartier for some $n \in \nn$. The singular category $\underline{\text{CM}}(A)$ is now rigid-free (as opposed to zero), which is the homological analog of smoothness. (An object $a$ in a triangulated category $\mathcal{T}$ with shift auto-equivalence $[\ ]$ is said to be rigid if $\Hom_\mathcal{T}(a,a[1])=0$; $\mathcal{T}$ is said to be rigid-free if every rigid object is isomorphic to the zero object.)} 

MMAs can also be used to construct a quiver in the case where $f$ does not factor at all. Indeed, for $cA_m$ isolated singularities, $f$ being irreducible is equivalent to the absence of a nontrivial (ordinary) crepant resolution $\tilde Y \dashrightarrow Y$, to the singular ring $R$ being $\qq$-factorial, with MMA given by $A \equiv  \End_R(M)$  (where $M$ is a so-called maximally modifying $R$-module -- as opposed to CM $R$-module, as is the case for NCCRs) \cite[Prop. 5.1]{iyama-wemyss-Qf}. The MMA is trivially obtained by presenting $R$ itself as a quiver with relations (see e.g.~the bottom quiver in the MMA ``hierarchy'' of \cite[Sec.~5.1]{iyama-wemyss-Qf}). The arrows are the generators of the polynomial ring $\cc[u,v,z,t]$, subject to the hypersurface equation and the commutativity relations (e.g. $uv=vu$, and so on). However these relations cannot be integrated to a superpotential (given there are more relations than arrows).\footnote{We would like to thank M.~Wemyss for discussions on this point.} 

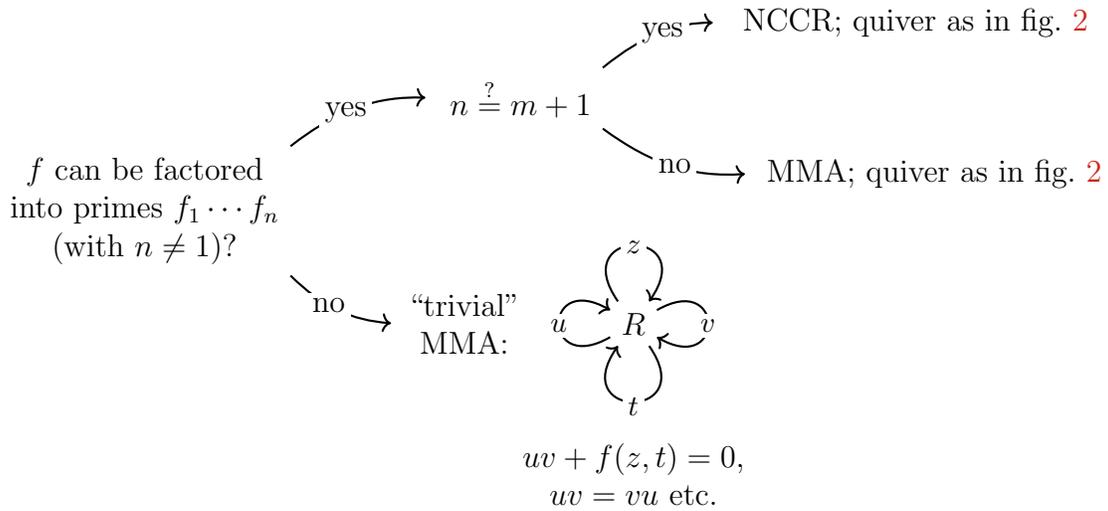
\begin{figure}[!ht]
    \centering
    \begin{tikzpicture}[->,thick, scale=1]
\node (Q1) at (0,0) [align=center] {$f$ can be factored \\ into primes $f_1 \cdots f_n$ \\ (with $n \neq 1$)?};

\node (A1y) at (5,1.5) {$n \overset{?}{\equal} m+1$};
\node (A1n) at (6.5,-1.5) {$R$};

\path[every node/.style={font=\sffamily\small, fill=white,inner sep=2pt}]
(A1n) edge [loop, out=210, in=150, looseness=7.5,pos=0.5] node[below=-4pt] {$u$}(A1n) 
(A1n) edge [loop, out=210+180, in=150+180, looseness=7.5,pos=0.5] node[below=-4pt] {$v$}(A1n) 
(A1n) edge [loop, in=60-5, out=120+5, looseness=7.5,pos=0.5] node[below=-4pt] {$z$}(A1n) 
(A1n) edge [loop, in=60-5+180, out=120+5+180, looseness=7.5,pos=0.5] node[below=-4pt] {$t$}(A1n) ;

\node (mma) at (4.25,-1.5) [align=center]{``trivial'' \\ MMA:};
\node (rel) at (6.5,-3.5) [align=center]{$uv+f(z,t)=0$, \\ $uv=vu$ etc.};

\draw [->,bend left=20,looseness=1,pos=0.4,shorten >= 5pt] (Q1.north east) to node[inner sep=2pt,fill=white]  {yes} (A1y.west);
\draw [->,bend right=20,looseness=1,pos=0.4,shorten >= 2pt] (Q1.south east) to node[inner sep=2pt,fill=white]  {no} (mma.west);

\node (A2y) at (10.25,2.5) [align=left]{NCCR; quiver as in fig.~\ref{fig:quivnccr}};
\node (A2n) at (10.5,.5) [align=left]{MMA; quiver as in fig.~\ref{fig:quivnccr}};

\draw [->,bend left=20,looseness=1,pos=0.5,shorten >= 7pt] (A1y.north east) to node[inner sep=2pt,fill=white]  {yes} (A2y.west);
\draw [->,bend right=20,looseness=1,pos=0.5,shorten >= 4pt] (A1y.south east) to node[inner sep=2pt,fill=white]  {no} (A2n.west);
\end{tikzpicture}
    \caption{The various possibilities given the $cA_m$ threefold $uv+f(z,t)=0$.}
        \label{fig:workflow}
\end{figure}

We summarize the various possibilities for singular threefolds of the form $uv+f(z,t)=0$ in the workflow \ref{fig:workflow}.
Given a $cA_m$ threefold singularity $uv+f=0$ with factored $f$, its quiver is given by figure \ref{fig:quivnccr} with notation as in \eqref{eq:NCCRmaxflag}. Each vertex corresponds to an $\SU(N)$ gauge group, each arrow between two vertices to a bifundamental chiral multiplet, each loop at a given vertex to an adjoint chiral multiplet. The F-terms of the superpotential are given by the abstract relations satisfied by the arrows, which are obtained as prescribed in section \ref{sub:iw}.

We note in passing that the single-D3 moduli space of the field theory discussed around (\ref{eq:single-D3}) can also be recovered from the quiver, via a geometric invariant theory procedure (see e.g. \cite[Sec.~4.1 \& 4.2]{collinucci-fazzi-valandro}).

As an example, we will now discuss the MMA for YY-II$(3,3)$,
\begin{equation}
		\label{eq:E7} uv+z^3+ z t^3=0 \ ,\\
\end{equation}
which is an $E_7$ threefold (in the terminology introduced below (\ref{eq:D4})). The quiver is again the one in figure \ref{fig:reid}. The superpotential reads:
\begin{equation}\label{eq:WE7}
W_{\text{YY-II}(3,3)} = \Tr \left( \frac{1}{4}e_0^{4}+  \frac{1}{4}e_1^{4} + e_0\left[(\alpha_1\beta_1)^{2} - \alpha_2\beta_2 \right] - e_1 \left[(\beta_1\alpha_1)^{2} - \beta_2\alpha_2\right]\right)\ .
\end{equation}
The superpotential constraint fixes the R-charges to be $R({e_i}) = \tfrac{1}{2}$, $R(\beta_1)=\frac34 - R({\alpha_1})$, $R({\beta_2}) = \tfrac{3}{2}-R(\alpha_2)$. This yields the following gauge coupling beta functions
\begin{equation}
\beta_R = \beta_{M_1} = -\frac{3}{8} N <0\ .
\end{equation}
Therefore the UV model is expected to flow, but the existence of an IR fixed point cannot be ascertained with certainty. (Said differently, if one assumes the existence of a fixed point and runs $a$-maximization, one finds the R-charges $R({\alpha_1})=R({\beta_1}) = \tfrac{3}{8}$, and $R({\alpha_2})=R({\beta_2}) = \tfrac{3}{4}$. However, for these values, the $a$ and $c$ central charges do not agree at large $N$, which is impossible for a superconformal quiver \cite{benvenuti-hanany}.)

This example demonstrates then that the presence of an MMA does not guarantee an SCFT, as expected. We have examined other MMAs (see footnote \ref{foot:mma}) with similar results.


\section{NCCRs for orbifolds} 
\label{sec:orbifolds}

In this appendix we show how NCCRs can be used to obtain quiver gauge theories for some orbifold theories. While this can be done in principle with the Douglas--Moore prescription \cite{douglas-moore}, the NCCR technique can sometimes make it easier to find the relations, and hence the superpotential. We warm up with the well-known example $\mathbb{C}^3/\mathbb{Z}_2 \times \mathbb{Z}_2$, and then consider a more complicated $\mathbb{C}^3/\Gamma$, with $\Gamma$ finite and non-abelian in $\mathrm{SL}(3,\mathbb{C})$. (Actually the quiver and superpotential have already been constructed for all finite $\Gamma <\SO(3)$ \cite{nolla-sekiya}.)

Notice that, for all finite $\Gamma$'s in $\mathrm{SL}(3,\mathbb{C})$ (which are classified \cite{fairbairn-fulton-klink}), a Calabi--Yau metric is guaranteed to exist on $\mathbb{C}^3/\Gamma$, since the orbifold respects the $\SU(3)$ special holonomy. (Indeed the existence of a SE metric on the link can be confirmed by checking K-stability.)

\subsection{\texorpdfstring{The $\cc^3 / \zz_2 \times \zz_2$ orbifold}{The C3/Z2xZ2 orbifold}}

The (orbifold-invariant) hypersurface equation is given in this case by 
\begin{equation}\label{eq:c3z2z2}
t^2+xyz=0\,.
\end{equation}
The threefold is again of $cD_4$ type, and is moreover toric. A K-stability analysis would just confirm that it is a Calabi--Yau threefold, as expected by the orbifold construction.

\begin{figure}[ht!]
\centering
\includegraphics[scale=.85]{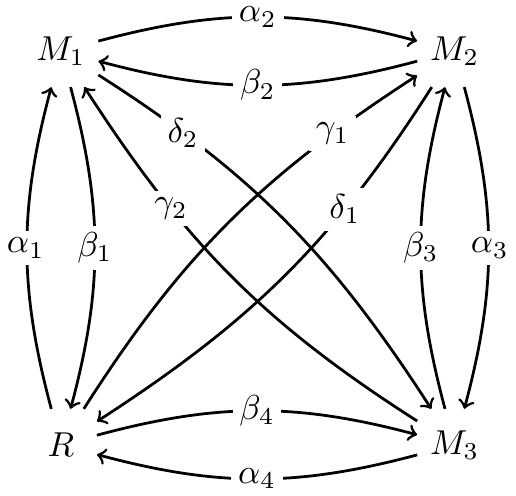}
\caption{The NCCR of $R=\cc[x,y,z,t]/\eqref{eq:c3z2z2}$.}
\label{fig:c3z2z2}
\end{figure}

The NCCR was constructed in \cite[Ex. 6.26]{iyama-wemyss-AR}, and has made an earlier appearance in physics in \cite[Sec. 5]{berenstein-leigh}. It is given by (\ref{eq:NCCR}) with three rank-one CMs:
\begin{equation}
\bigoplus_{i=1}^3 M_i \equiv (t,x) \oplus (t,y) \oplus (t,z)\,.
\end{equation}
These give rise to the familiar quiver in figure \ref{fig:c3z2z2}, with maps reading
\begin{equation}
\alpha_1=\beta_1 = \alpha_3=\beta_3 =x\,,\quad \alpha_2=\beta_2 = \alpha_4=\beta_4 =y\,,\quad \gamma_1=\delta_1 = \gamma_2=\delta_2 =z\,.
\end{equation}
The relations are generated by commutativity of these maps and can be integrated to the following superpotential:
\begin{multline}
W = \Tr \left(\beta_2\beta_1\gamma_1 -\gamma_1\alpha_3\alpha_4 +\alpha_3\gamma_2\alpha_2-\alpha_2\delta_1\alpha_1 \right. +\\ + \left. \delta_1\beta_4\beta_3-\beta_3\beta_2\delta_2-\beta_1\beta_4\gamma_2+\alpha_1\delta_2\alpha_4\right)\, .
\end{multline}

\subsection{\texorpdfstring{A non-abelian $\SL(3,\cc)$ orbifold}{A non-abelian SL(3,C) orbifold}}

\begin{figure}[ht!]
\centering
\includegraphics[scale=1]{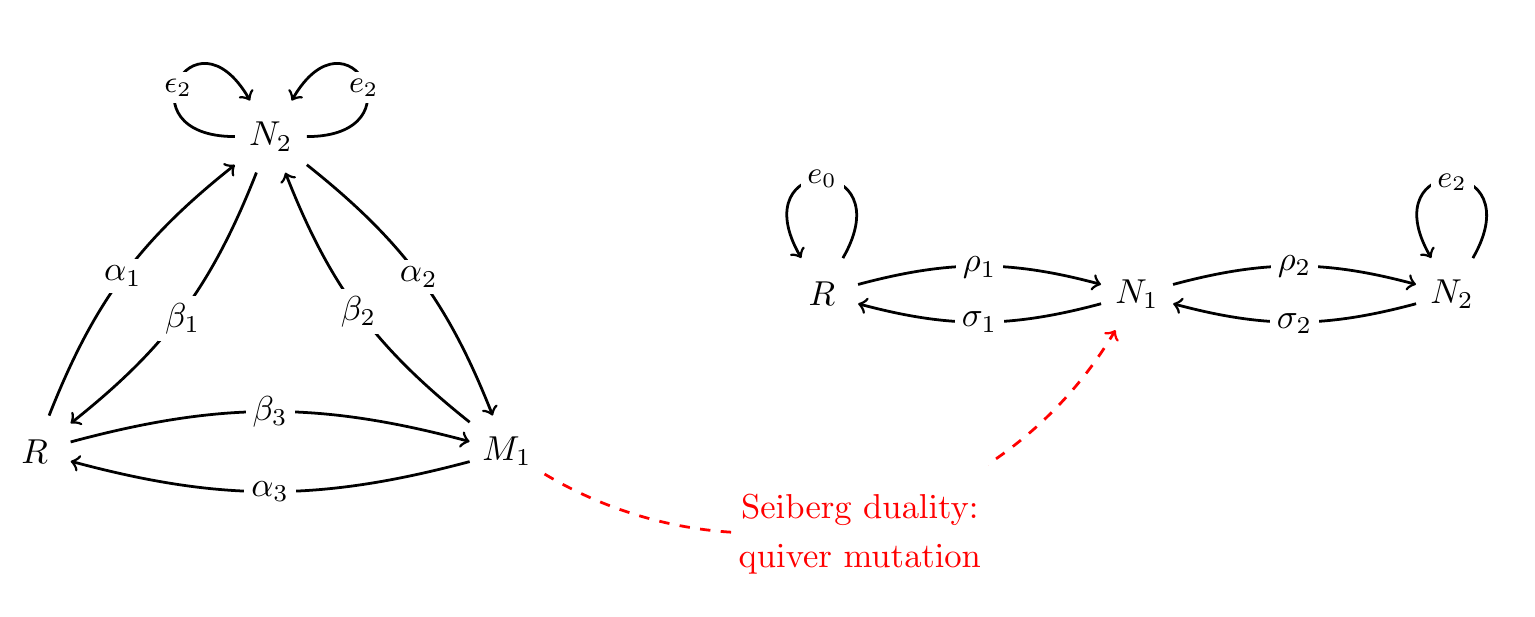}
\caption{The NCCR of $R=\cc[x,y,z,t]/\eqref{eq:D3}$ before and after Seiberg duality (i.e. categorical quiver mutation \cite{wemyss-hommmp}) performed at node $M_1$. The $M_i$ are rank-one CMs (and correspond to physical $\SU(N)$ groups), whereas the $N_i$ are rank-two (and correspond to $\SU(2N)$).}
\label{fig:orbifoldSL}
\end{figure}%
We now look at a more challenging example. Consider the orbifold of $\cc^3$ by 
\begin{equation}
\mathbb{D}_{2\cdot 3} = \left\langle \diag(\epsilon,\epsilon^2,1)\ ,\ \begin{pmatrix} 0 &1&0\\1&0&0\\0&0&-1 \end{pmatrix}\right\rangle < \SL(3,\cc)\ ; \quad \epsilon^3=1\ .
\end{equation}
This is the dihedral group of order six. The (orbifold-invariant) hypersurface equation is given by 
\begin{equation}\label{eq:D3}
t^2-z(x^2-4y^3)=0\,.
\end{equation}
The threefold is again of $cD_4$ type, as can easily be verified. The NCCR \cite[Ex. 7.7]{wemyss-hommmp} gives rise to the quiver in figure \ref{fig:orbifoldSL}, where as a curiosity we have also added a Seiberg-dual phase. The superpotential $W$ ($W'$) in the left (right) frame of figure \ref{fig:orbifoldSL}, i.e.~before (after) Seiberg duality, is given by
\begin{subequations}
\begin{align}
W &= \Tr\left(\beta_1\beta_3\beta_2+\alpha_2\alpha_3\alpha_1 - \epsilon_2^2 \beta_1\alpha_1 - \epsilon_2^2 \alpha_2\beta_2+2\epsilon_2 e_2^2\right)\, ,\\
W' &=\Tr\left(-\sigma_1 e_0 \rho_1 - \rho_2 e_2^2 \sigma_2 + \rho_2\sigma_2\rho_2\sigma_2 \sigma_1\rho_1\right)\, .
\end{align}
\end{subequations}
The maps can be worked out by looking at the ideals defining the CMs (see \cite[Sec. 3.3]{nolla-sekiya}). For example, the rank-one CM $M_1 = (t,z)$ is associated to the MF $(\phi,\psi)_2$ with $\psi=\left[\begin{smallmatrix} t & z \\ x^2-4y^3 & t \end{smallmatrix} \right]$ (i.e. $M_1 = \coker \psi$).



\bibliography{mpa}
\bibliographystyle{at}

\end{document}